\documentclass[twocolumn,superscriptaddress,aps,prr,longbibliography]{revtex4-2}
\usepackage[dvips]{epsfig}
\usepackage[dvips]{graphicx}
\usepackage{latexsym,amsmath,verbatim}
\usepackage[usenames,dvipsnames]{color}
\usepackage{enumitem}
\usepackage{amssymb}
\usepackage{graphicx}   
\usepackage{subfigure}
\usepackage{float}
\usepackage[normalem]{ulem}

\newcommand{\new}[1]{{\color{black}{#1}}}
\newcommand{\newr}[1]{{\color{black}{#1}}}

\DeclareMathOperator{\tr}{tr}
\DeclareMathOperator{\cv}{CV}
\DeclareMathOperator{\isi}{ISI}

\begin{document}     

\title{Hybrid-type synchronization transitions: where incipient oscillations,
  scale-free avalanches, and bistability live together}

\author{Victor Buend{\' i}a}
\affiliation{Departamento de Electromagnetismo y F{\'i}sica de la Materia
  e Instituto Carlos I de F{\'i}sica Te{\'o}rica y
  Computacional. Universidad de Granada, E-18071 Granada, Spain}
\affiliation{Dipartimento di Matematica, Fisica e Informatica,  Universit\`a di Parma, via G.P. Usberti, 7/A - 43124, Parma, Italy}
\affiliation{INFN, Gruppo Collegato di Parma, via G.P. Usberti, 7/A -
  43124, Parma, Italy} 
\author{Pablo Villegas}
\affiliation{Networks Unit, IMT School for Advanced Studies, Lucca, Italy \& Instituto Carlos I de F{\'i}sica Te{\'o}rica y
  Computacional. Universidad de Granada, E-18071 Granada, Spain}
\author{Raffaella Burioni}
\affiliation{Dipartimento di Matematica, Fisica e Informatica,
  Universit\`a di Parma, via G.P. Usberti, 7/A - 43124, Parma, Italy}
\affiliation{INFN, Gruppo Collegato di Parma, via G.P. Usberti, 7/A -
  43124, Parma, Italy} 
\author{Miguel A. Mu\~noz}
\affiliation{Departamento de Electromagnetismo y F{\'i}sica de la Materia
  e Instituto Carlos I de F{\'i}sica Te{\'o}rica y
  Computacional. Universidad de Granada, E-18071 Granada, Spain}

 \date{\today}

\begin{abstract}
  The human cortex is never at rest but in a state of sparse and noisy neural activity that can be detected at broadly diverse resolution scales. It has been conjectured that such a state is best described as a critical dynamical process ---whose nature is still not fully understood--- where scale-free avalanches of activity emerge at the edge of a phase transition.  In particular, some works suggest that this is most likely a synchronization transition, separating synchronous from asynchronous phases.  \newr{Here, by investigating a simplified model of coupled excitable oscillators} describing the cortex dynamics at a mesoscopic level, we investigate the possible nature of such a synchronization phase transition. \newr{Within our modelling approach,  we conclude that ---in order to reproduce all key empirical observations, such as scale-free avalanches and bistability, on which fundamental functional advantages rely---} the transition to collective oscillatory behavior needs to be of an unconventional hybrid type, with mixed features of type-I and type-II excitability, opening the possibility for a particularly rich dynamical repertoire. 
  \end{abstract}

\maketitle

\section{Introduction}

Neurons in the cerebral cortex fire in a rather irregular and sparse way, even in the absence of external stimuli or tasks \cite{Softky,Arieli,Abeles}.  This activity is also manifested at large scales in the so-called \emph{resting-state networks} \cite{Fox,Beckmann}.  Understanding the origin and functionality of such an energetically-costly fluctuating ``ground state'' is a fundamental question in neuroscience, essential to shed light on \new{how the cortex processes and transmits information} \cite{Latham,Mattia-Vives,Miller,dynamicbrain,Breakspear-REVIEW}.

Two twin sides of spontaneous neuronal activity are \emph{synchronization and avalanches}. Depending mostly on cortical regions and functional states, diverse synchronization levels across a continuum spectrum are observed. Both synchronous and asynchronous states are retained to be essential for diverse aspects of information processing; e.g., neuronal synchronization is at the root of collective oscillatory rhythms, a crucial aspect for information transmission between distant areas \cite{Buzsaki,Terry-waves,Deco-synchro}, while asynchronous states also play key roles for information coding \cite{Renart}.  Accumulating evidence --including results from the Human Brain Project \cite{HBP}-- suggests that the ground state of spontaneous activity of a healthy cortex lies close to the edge of a synchronization phase transition, neither exceedingly synchronous nor overly incoherent, allowing for transient and flexible levels of neural coherence, as well as a very-rich dynamical repertoire \cite{HBP,Cabral-review, Breakspear-REVIEW}. Indeed, abnormalities in the synchronization level are linked to pathologies such as e.g. \new{Parkinsonian disease (excess) and autism (deficit) \cite{Parkinson,autism,Kandel-book}.}
 
On the other hand, neuronal activity is also observed to propagate in the form of irregular outbursts, termed \emph{neuronal avalanches} \cite{BP2003,Petermann,Plenz-2}. These are cascades of activations clustered in time and interspersed by periods of relative quiescence, which \new{are robustly observed across cortical areas, species, and resolution scales} \cite{Schuster,Breakspear-REVIEW,Chialvo2010,Plenz2021}.  Neural avalanche sizes ($S$) and durations ($T$) are empirically observed to be scale-free, i.e., their associated probability distributions exhibit power-law tails $P(S)\sim S^{-\tau}$, $P(T)\sim T^{-\alpha}$, and the mean avalanche size obeys $\langle S(T)\rangle\sim T^{\gamma}$ with $\gamma$ fulfilling the scaling relation $\gamma=(\alpha-1)/(\tau-1)$, which is a fingerprint of criticality \cite{Avalanches,Friedman2012}.  \new{In other words, neuronal avalanches are scale-free, exhibiting signatures of dynamical criticality \cite{BP2003}, with critical exponents close to those of a critical branching process, describing marginal propagation of activity \cite{BP2003,Petermann,Plenz-2,Serena-BP}.}

These empirical observations triggered the development of the ``\emph{criticality hypothesis}'', conjecturing that the cortex might extract crucial functional advantages --e.g., large sensitivity and optimal computational capabilities-- by operating close to a critical point \cite{BP2003,Schuster,Chialvo2010,Mora-Bialek,RMP,Zierenberg,WiltingReverberating}. However, there is still no consensus on what type of phase transition is required for such a critical behavior; \new{ a critical branching process, \new{as mentioned above, is the most common and straightforward interpretation}, but alternative scenarios have been put forward} \cite{Deco-whole-brain,Copelli-PRL,Copelli-Porta,Friedman2012,Orlandi-2018}.  Among these, a Landau-Ginzburg theory has been recently proposed to describe cortical dynamics at a mesoscopic level suggesting that the cortex might operate in a regime close to the edge of a synchronization phase transition \new{at which, quite remarkably, scale-free avalanches emerge in concomitance with incipient oscillations \cite{LG}. This scenario that had been previously suggested in the literature \cite{Poil,Zhou-2017,LG,Copelli-Porta,Copelli-2019,Zhou-2020,PittorinoChaos} and that is supported by empirical evidence \cite{Gireesh,Yang,Miller-Plenz,HBP}}. In spite of these advances, a minimal model ---simpler than the analytically un-tractable Landau-Ginzburg theory--- capturing the gist \new{of such a synchronization phase transition and allowing for in-depth theoretical understanding is still missing.}

Here we pose the following questions: can scale-free avalanches possibly occur at the critical point of the canonical model for phase synchronization? If not, which is the minimal model for synchronization able to accommodate them?  What type of phase transition does it exhibit? \new{Answering these questions will pave the road for the more ambitious goal of constructing a statistical mechanics of cortical networks, shedding light on the collective states --with crucial functional roles-- that they sustain, as well as advancing our general understanding of phase transitions.}

\section{Lack of scale-free avalanches in the annealed Kuramoto model }

\begin{figure*}[hbtp]
  \centering
  \includegraphics[width=1.7\columnwidth]{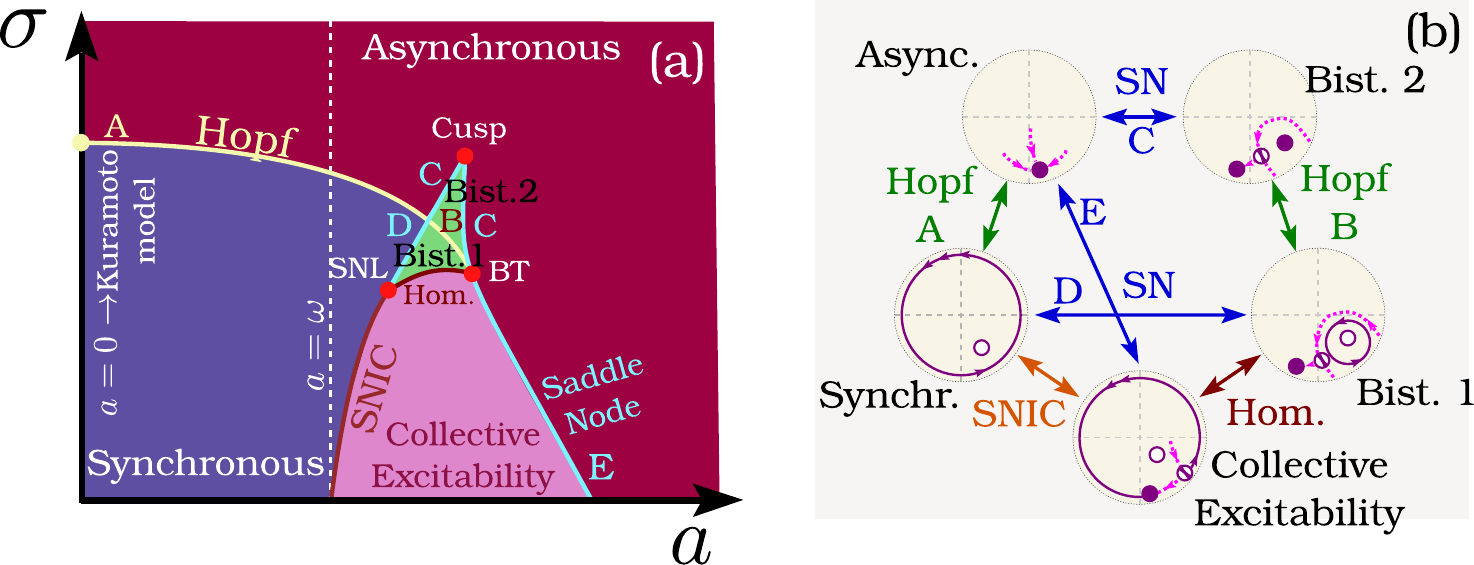}
  \caption{Illustrative diagrams sketching in a qualitative way the phase diagram ---underlining the main bifurcations and phases---  as analytically obtained. It reveals the existence of synchronous and asynchronous states, and a collectively-excitable regime, separated by diverse types of bifurcations for the collective order parameter. The central triangular-shaped region \new{(whose size has been increased for clarity)} describes a bistability regime; its vertices correspond to codimension-2 bifurcations \new{(see main text)}. Finally, there is also a homoclinic line (Hom) linking the SNL and the BT points. \textbf{(b)} Sketch of the different phases/regimes represented in terms of their corresponding complex Kuramoto order parameter, $Z$; a point represents each regime in the unit circle (since $|Z|\leq 1$) with filled circles describing fixed points, open circles stand for unstable fixed points with associated limit cycles, and mixed-color circles describe saddles. Bifurcations between different regimes are indicated as arrows.}
  \label{sketch}
\end{figure*}

We start by analyzing the canonical model for phase synchronization, customarily used in neuroscience \cite{Breakspear-Kuramoto, Cabral,Stoop,Villegas1} and other fields \cite{Pikovsky-book}: the Kuramoto model \cite{Kuramoto,Kuramoto-book,Acebron,Pikovsky-book}. \new{However, given that ---as revealed by the Landau-Ginzburg model \cite{LG}--- node heterogeneity is not an essential ingredient to generate avalanches, and that, on the other hand, noise is inherent to neural dynamics, we consider here the annealed version of the Kuramoto model with homogeneous frequencies} \cite{Pikovsky-book,Villegas1}, i.e.
 \begin{equation}
\dot{\varphi_j}(t)=  \omega + \dfrac{J}{N}\sum_{i=0} ^N \sin\left( \varphi_i(t) - \varphi_j(t) \right) + \sigma \eta_j(t)
\label{Kuramoto}
\end{equation}
where the phase $\varphi_j(t)$ describes the dynamical state of the $j$-th oscillatory unit, with $j \in [1,N]$, $\omega$ is a common intrinsic frequency, $\eta_j(t)$ a zero-mean unit-variance Gaussian white noise with amplitude $\sigma$, and $J$ is the coupling strength with all the neighbours on, e.g., a fully-connected network \cite{Kuramoto,Kuramoto-book,Strogatz-Kuramoto,Pikovsky-book, Acebron}. \new{As is well known, Eq.(\ref{Kuramoto})} exhibits a synchronization phase transition where the synchronization (Kuramoto) order parameter, $Z= \langle e^{i \varphi} \rangle$, experiences a (supercritical) Hopf bifurcation from a fixed point to a limit cycle, revealing the emergence of collective oscillations \cite{Kuramoto,Acebron}.

\new{ In what follows, we set to analyze if this model (and related models) can exhibit scale-free outbursts of activity. For this purpose, we define and measure avalanches following the standard protocol in neuroscience} \cite{BP2003,Schuster}, which relies on the identification of individual-unit ``spikes'' \new{(in the case of models consisting of oscillators ``spikes'' can be identified in an effective way as crossings over a given phase value)} as well as on the definition of a time discretization window, needed to cluster close-in-time spikes together \cite{BP2003} \new{(details of the protocol are carefully explained in Appendix A; see also \cite{Martinello})}.  Extensive computational simulations (reported in detail in Appendix B) reveal that neither at the critical point of Eq.(\ref{Kuramoto}) nor around it \new{scale-free avalanches can be found; i.e. $P(S)$ and $P(T)$ always show exponential decays, even if other standard quantities are known to exhibit scaling for different versions of the Kuramoto model \cite{Hong,Odor1,Odor2}}.

\section{Hybrid-type synchronization emerges from excitable oscillators}

To search for a better suited \new{minimal model for synchronization with scale-free avalanches}, we scrutinize the Landau-Ginzburg theory (LG) in \cite{LG}, \new{known to exhibit these two features in concomitance}.  In a nutshell, the LG model consists of a set of diffusively coupled units, each of which represents a mesoscopic region of the cortex, and is described by a set of two dynamical equations for the local density of: (i) neuronal activity and (ii) available synaptic resources, respectively  \new{(see Appendix C for a more detailed presentation of the model).}
A salient feature is that as the control parameter \new{(which can be taken to be the maximum possible level  of synaptic resources)} is increased, \new{each isolated individual unit} can experience an \emph{infinite-period bifurcation} from a low-activity fixed point to a limit cycle ---with zero-frequency and fixed-amplitude at the bifurcation point--- where both activity and synaptic resources oscillate in a ``spike-like'' manner, \new{i.e., with a phase-dependent angular velocity, in contrast with the sinusoidal oscillation in the Kuramoto model of Eq.\ref{Kuramoto}, 
suggesting that a different minimal model is needed to describe individual units.}
Moreover, isolated  units can produce spikes even when they are slightly below threshold owing to the effect of noise; in other words, they behave as \emph{type-I excitable units} \cite{Izhikevich-book,Ojalvo-review} (see Appendix D for a brief account of excitability types). Importantly, \new{for sufficiently large values of the control parameter,
 the coupled oscillatory units
become synchronized and, at the edge of the synchronization phase transition, scale-free avalanches emerge \cite{LG}.
Finally, let us recall that a similar phenomenology is obtained in a variant of the LG model, including inhibitory neural populations \cite{LG}, a well-recognized key player in the generation of neural rhythms \cite{Traub}.}

Guided by these observations, we consider a set of coupled oscillatory units, each of them represented by the canonical form of type-I excitable units (see Appendix D) or ``active rotors'' as defined by 
\begin{equation}
\dot{\varphi}=\omega + a\sin\varphi,
\end{equation} where $\omega$ and $a$ are parameters \cite{Strogatz-book,Adler}.  For $a>\omega$ the deterministic dynamics of each isolated unit exhibits a stable fixed point at $\varphi^* = -\arcsin(\omega/a)$, as well as a saddle point with the opposite sign.  The addition of a stochastic term $\sigma \eta(t)$ can induce fluctuations beyond the saddle, thus generating large excursions of the phase before relaxing back to its equilibrium. \new{On the other hand}, for $a<\omega$ the system oscillates with phase-dependent angular velocity and, as the ``saddle-node into invariant circle'' (SNIC) bifurcation point $a_c=\omega$ is approached, the frequency of the oscillations vanishes, implying that the period becomes infinite, while the amplitude remains constant, as in the LG model (let us recall that type-I excitability can rely either on a ``saddle-node into invariant circle'' (SNIC) bifurcation, as in the equation above, or in a homoclinic bifurcation, as in the LG model \cite{LG}: both types share the common relevant feature of generating spike-like infinite-period oscillations at the bifurcation point \cite{Izhikevich-book}).

Thus, finally, the full model reads
\begin{equation} 
\dot{\varphi_j}=\omega + a\sin\varphi_j + \dfrac{J}{M_j}{\sum_{i \in n.n.j} ^{M{j}}} \sin\left( \varphi_i - \varphi_j \right) + \sigma \eta_j(t), \label{SNIC-coupled} 
\end{equation}
where the sum runs over the ($M_j$) nearest neighbors of unit $j\in 1,2,...N$ in a given network. We consider versions of the model embedded on fully-connected (FC) networks ($M_j=N, \forall j$), which are useful for analytical approaches and on two-dimensional (2D) lattices (as, at large scales, the cortex can be treated as a 2D sheet \cite{Breakspear-REVIEW,LG}).
\new{Let us remark that} Eq.(\ref{SNIC-coupled}) is sometimes called Shinomoto-Kuramoto model or Winfree's ring model \cite{SK-1,Winfree} and has been analyzed in diverse contexts \cite{SK-1,SSK, Geier-gaussian, Pikovsky-cumulants}.  Its collective state can be quantified by the Kuramoto-Daido parameters:
 \begin{equation}
Z_k = \langle e^{ik\varphi} \rangle \equiv \frac{1}{N} \sum_{j=1}^N {e^{ik\varphi_j}} \equiv R_k e^{i\psi_k}, 
\end{equation}
where $ k$ is any positive integer number;
for $k=1$ this coincides with the usual Kuramoto parameter.
Analytical progress is achieved in a mean-field approximation \new{(which, as usual, becomes exact for infinitely large FC networks) which consists in:} (i) writing down a continuity (Fokker-Planck) equation for the probability distribution of phase values, $P(\varphi, t)$ \cite{SK-1,SSK,Geier-gaussian, Pikovsky-cumulants,Acebron}, (ii) 
expanding $P(\varphi, t)$ in power series, and (iii) writing an infinite hierarchy of coupled equations for its coefficients (which coincide with the $Z_k$'s; see Appendix E and \cite{SK-1,Geier-gaussian}):
\begin{eqnarray}
\dot Z_k &=& Z_k(i \omega k - \frac{k^2 \sigma^2}{2} ) + \frac{a k}{2} \left( Z_{k+1} - Z_{k-1} \right) \nonumber \\ &+& \frac{J k}{2} \left( Z_1 Z_{k-1} - \bar Z_1 Z_{k+1} \right).
\label{zk_system}
\end{eqnarray}
where the bar stands for complex conjugate. The associated phase diagram has been scrutinized in the literature by using different low-dimensional closures for this infinite hierarchy of coupled equations. For instance, using the Ott-Antonsen (OA) ansatz \cite{OA} or other more-refined closures \cite{Geier-gaussian, Pikovsky-cumulants}, one can obtain the phase diagram, summarized in Figure \ref{phase_diagram} \new{(a careful discussion can be found in Appendices 6 and 7)}.  \new{Inspection of the phase diagram reveals} that there are two main types of collective dynamical regimes: oscillations (synchronous states) and stable fixed points (\new{corresponding to either high-activity asynchronous states in the upper part of the diagram or low-activity states in the lower/right part}). These are separated by different types of bifurcation lines. In particular, for low-noise amplitudes, $\sigma$, as the control parameter $a$ is increased there is a collective SNIC bifurcation from the oscillatory regime to a phase characterized by a stable fixed point with very-low spiking activity, but susceptible to collectively react to external inputs, called the \emph{collective-excitability} phase.  In analogy with the classification of excitability types we refer to this as \emph{type-I synchronization transition}.  Conversely, for small values of $a$, by increasing $\sigma$ one encounters a collective Hopf bifurcation, a \emph{type-II synchronization transition}, to a high-activity asynchronous state (see Appendix D for a discussion of excitability types). \new{Thus, there are two main lines of bifurcations to a collectively oscillatory phase in the two-dimensional phase diagram: one of type-I and one of type-II}.

\new{Remarkably, the above two bifurcations lines} cannot possibly intersect each other owing to topological reasons \cite{Izhikevich-book}, so there is not such a thing as a ``tricritical'' point. \new{Instead, in the region where they come close to each other, there exists a triangular-shaped region of bistability (green area in Figure \ref{sketch}a and b) \cite{SSK, Strogatz-childs} delimited
  by three bifurcation lines and three co-dimension-2 bifurcations (\new{which are signaled by red circles in Figure \ref{sketch}b)}.

\begin{figure}[hbtp]
  \centering
  \includegraphics[width=0.85\columnwidth]{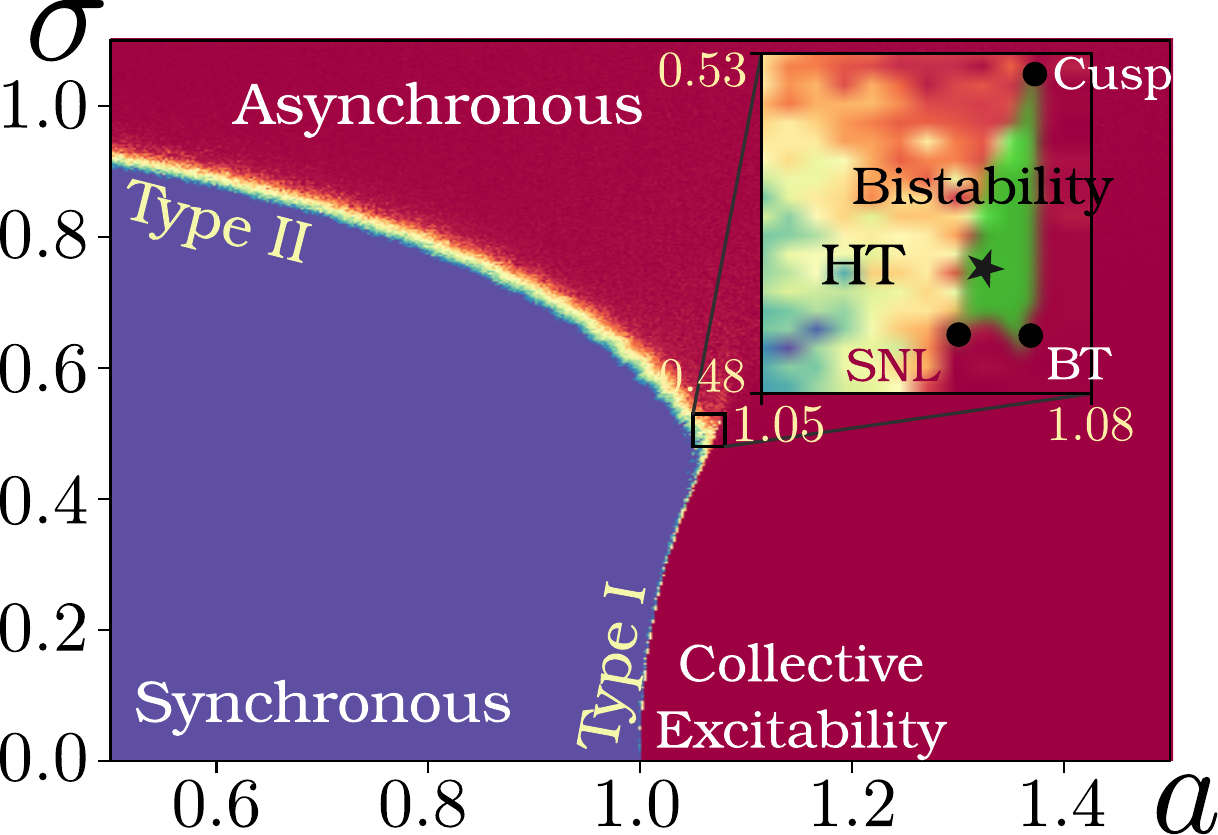}
  \caption{\textbf{Phase diagram and bifurcations of Eq.(\ref{SNIC-coupled}) on a fully-connected network.}  \textbf{(a)} Phase diagram computed using direct simulations on a fully-connected network with $N=5000$ oscillators ($\omega=1$, $J=1$). Collective oscillations are computationally detected with the Shinomoto-Kuramoto order parameter (see Appendix G) \cite{SK-1}. The  location of the bistability region was established by numerically solving Eq.(\ref{zk_system}) for the first $k=50$ harmonics ($Z_{51}=0$). The inset shows a zoom of the bistability region, where codimension-2 points are marked;  a star indicates a point with scale-free avalanches. The three segments (red, green and blue) indicate three possible types of transition to synchronization.}
  \label{phase_diagram}
\end{figure}

\new{For the sake of completeness and illustration}, Figure \ref{sketch}b presents a sketch illustrating the (complex) order parameter in the different phases and the different bifurcation lines. In particular, there is a} \emph{Bognadov-Takens} (BT) point, where the Hopf-bifurcation line finishes, colliding tangentially with a line of saddle-node bifurcations; a \emph{saddle-node-loop} (SNL) where the line of SNIC bifurcations ends, becoming a standard saddle-node line; and a \emph{cusp}, where two saddle-node bifurcation lines collide.  Observe that the bistability region is divided into two halves by the Hopf-bifurcation line, so the regime of collective excitability coexists with either oscillations below the Hopf line or the high-activity asynchronous state above the Hopf line. In other words, in this region, the Hopf bifurcation occurs in one of the branches of two coexisting solutions, i.e., in concomitance with bistability.

\begin{figure*}[hbtp] \begin{center}
  \includegraphics[width=1.5\columnwidth]{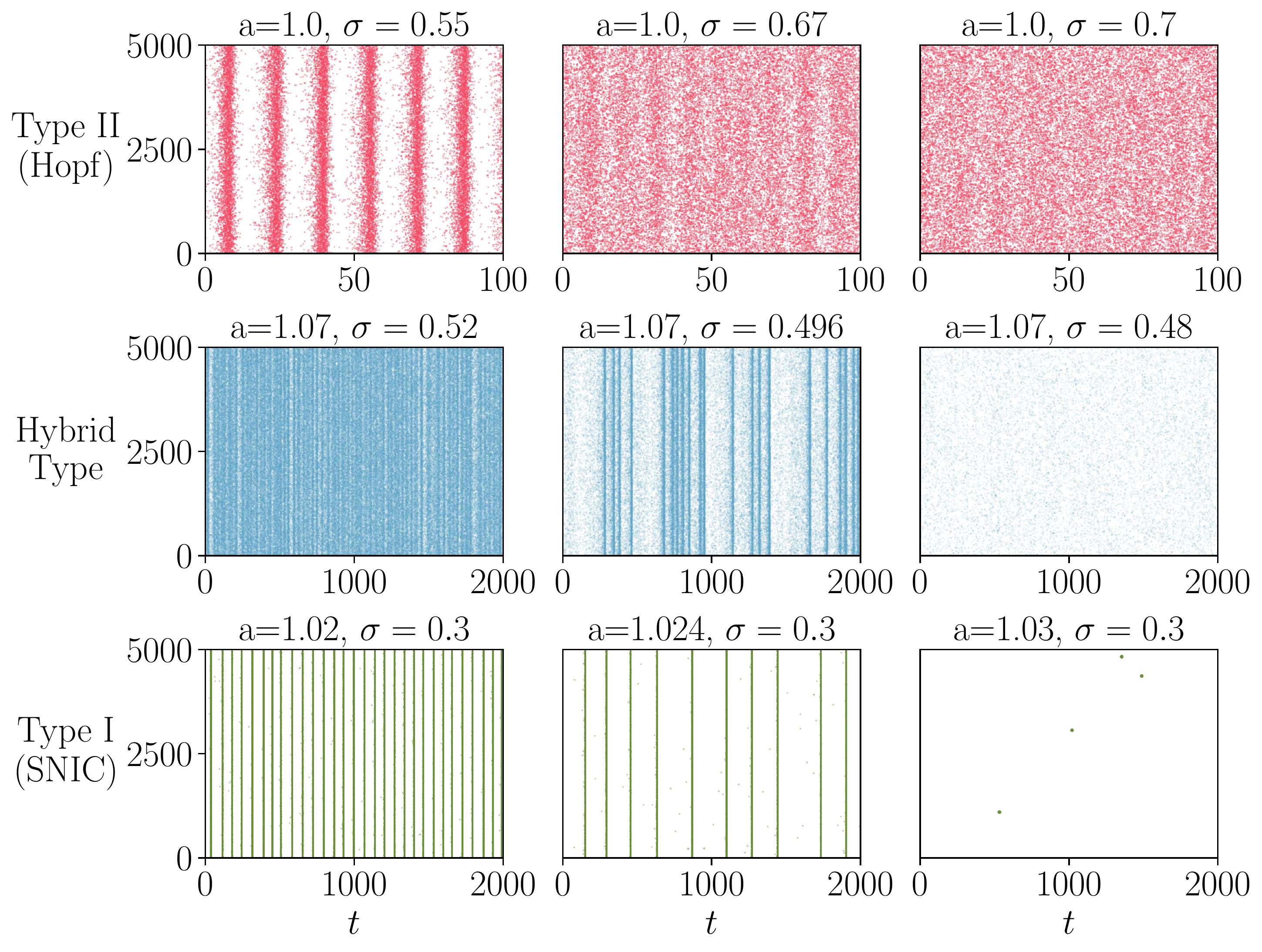}
  \caption{Raster plots in a fully-connected network of size $N=5000$ for each of the three considered cases (as indicated in Figure \ref{phase_diagram}) --- Hopf (type-II) bifurcation, hybrid-type (HT) synchronization, and SNIC(type-I) bifurcation (from top to bottom)--- in the synchronous phase (left column), right at the transition point or very slightly within the synchronous/oscillatory phase (central column), and in the asynchronous phase (right column).}\label{PanelMF}
  \par\end{center}
\end{figure*}

\begin{figure*}[hbtp]
 \begin{center}
    \includegraphics[width=1.90\columnwidth]{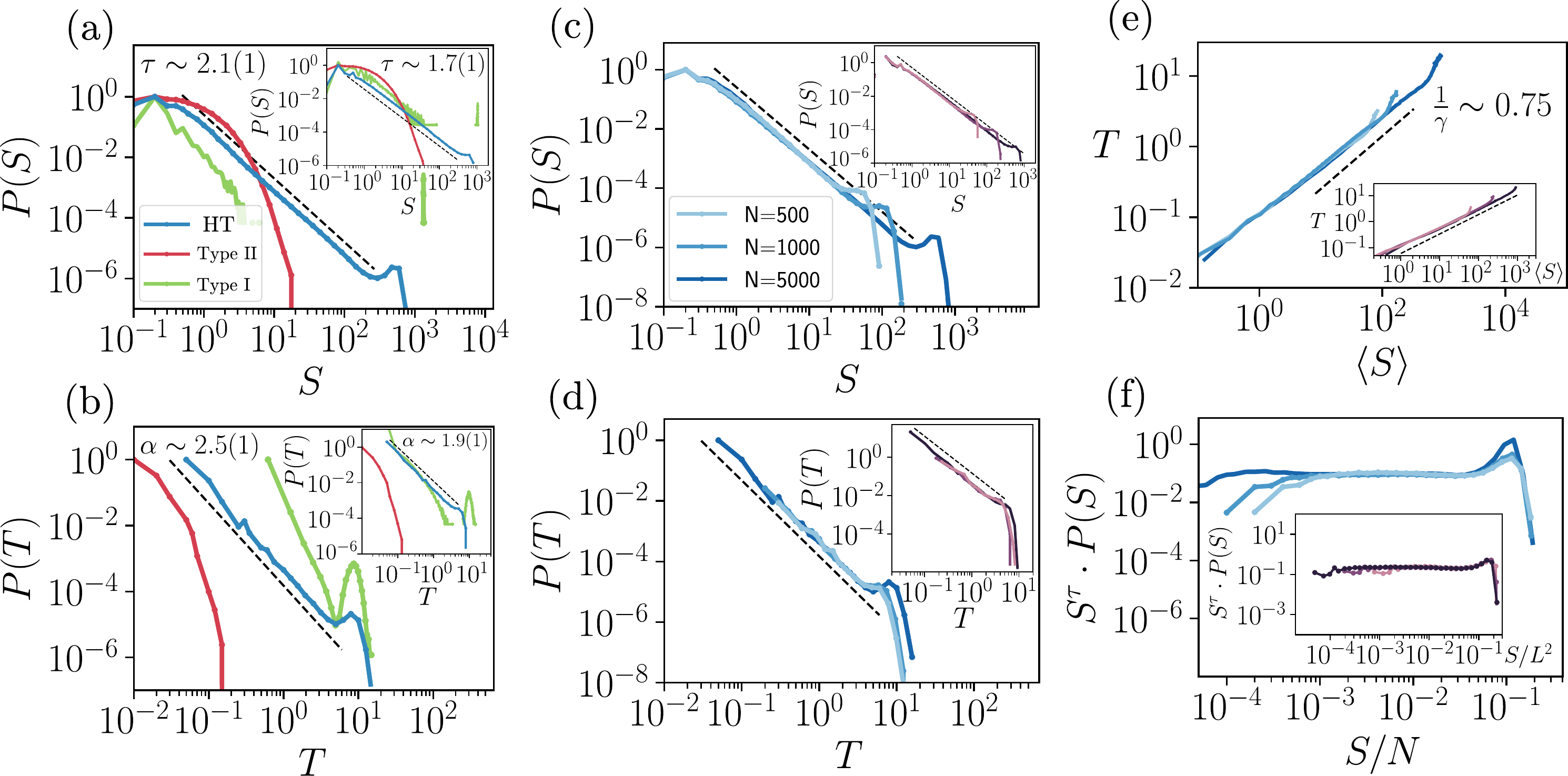}
    \caption{ \textbf{Avalanche distributions.} \textbf{(a,b)} Avalanche-size and duration distributions for three different types of synchronization transitions (right at the transition points): type-I (SNIC) transition (green lines, $a = 1.  024$, $\sigma = 0.3$); type-II (Hopf) transition  (red lines, $a = 1.04$, $\sigma = 0.575$), and HT transition (blue lines, $a = 1.07$, $\sigma = 0.499$)
in FC networks ($N=5000$). Only the last one exhibits clear cut power-law behavior, both for size and duration distributions. \textbf{Insets}: As in the main Figure but for simulations in a 2D lattice (size $64^2$): type-I transition ($a = 0.99$, $\sigma = 0.05$), type-II transition ($a = 0.60$, $\sigma = 0.64$), and HT synchronization transition ($a = 0.98$, $\sigma = 0.185$). \textbf{(c,d)} Finite-size scaling analysis of $P(S)$ and $P(T)$ in FC networks of different sizes (as specified in the legend) in the HT regime. \textbf{Insets}: As in the main Figures but for 2D lattices of sizes ($N=16^2$, $N=32^2$, $N=64^2$). \new{ \textbf{(e)} Averaged avalanche size as a function of the duration for different system sizes. \textbf{Inset}: As in the main Figure, but for simulations in a 2D lattice for different system sizes. \textbf{(f)} Distribution of avalanche sizes multiplied by $s^{\tau}$ to obtain an asymptotically flat curve (to ease visual inspection of scaling) and re-scaled as a function of the system size to collapse the different curves. As expected from true scale invariance, the distributions become flat and collapse for different system sizes after rescaling. \textbf{Parameters at the HT transition} $(a,\sigma)$ for fully connected  networks, $N=500:~(1.07,0.520)$, $N=1000:~(1.07,0.505)$, $N=5000:~(1.07,0.496)$ and 2D networks, $L=16:~(0.995,0.192)$, $L=32:~(0.982,0.190)$, $L=64:~(0.98,0.185)$, $J=\omega=1$. Dashed lines are guides to the eye showing the value of the fitted exponents.}
}
     \label{avalanches}
\end{center}
\end{figure*}

In order to obtain a highly-accurate phase diagram needed for forthcoming analyses, we complemented the above analytical approximations (i.e., closures) with extensive computational analyses of the complete stochastic system (size $N = 10^4$) as well as a direct numerical-integration of Eq.(\ref{zk_system}) truncated at sufficiently large values of $k$ ($k=50$).  Results of this combined approach are summarized in Figure \ref{sketch}a which shows that the overall shape of the phase diagram is qualitatively identical to the one predicted by low-dimensional closures, \new{but with a slightly smaller bistability region}. \new{Remarkably, a very similar phase diagram, with the same type of phases and phase transitions is found in 2D lattices (see Appendix H).}

\newr{ The reader can gain insight into the dynamics in each phase and around the different transition points by inspecting Figure 3, as well as in Supplementary Material 1 \cite{SI}  and contains a set of videos for diverse parameter choices.}

The moral of these findings is that, to analyze \new{general synchronization transitions}, \new{it is necessary to consider} not only the standard type-I and type-II cases, but also more complex scenarios, including the case in which the transition to synchrony occurs in concomitance with bistability, i.e., when incipient oscillations coexist with low-activity asynchronous states; we call these \emph{``hybrid-type (HT) synchronization transitions''}.  \new{This is illustrated in Figure\ref{PanelMF} (obtained from simulations on FC networks), that shows representative raster plots around each of these
  possible transition types} (colored segments in Figure \ref{sketch}a identify these three examples in the phase diagram):
  type-I (SNIC), type-II (Hopf),  and HT synchronization transition, respectively.
  \new{In particular, it shows results slightly within the synchronous phase (left), very close to criticality (central), and in the asynchronous phase (right).
    Surprisingly, even naked-eye inspection already reveals that raster plots nearby the HT transition exhibit a much larger dynamical richness (see below for a more quantitative analysis).}
 
\section{Scale-free avalanches near the hybrid-type synchronization transition}
Following the protocol for avalanche detection (Appendix A) we determined the statistics of avalanche sizes and durations in these three scenarios. As shown in Figure \ref{avalanches} for FC networks, power-law distributed avalanches do not emerge at the  type-II (Hopf) transition \new{(red curves)} nor at the type-I (SNIC) \new{(green curves)}. For instance,
   for the Hopf-bifurcation case results resemble, not surprisingly, those for the annealed Kuramoto model which also exhibits a type-II collective bifurcation,  while for the type-I case
 there are large collective events recruiting most of the units in the system into anomalously-large avalanches, separated by extremely long periods of quiescence (as illustrated in the raster plot in Figure \ref{PanelMF}).  In any case, we observe no signature whatsoever of scale-free avalanches at any point along the line of SNIC bifurcations (green dots in Figure \ref{avalanches}; see also Appendix G).

On the other hand, when entering the synchronous phase through the HT transition, clean scale-invariant avalanche distributions are observed at criticality. \new{These avalanches} span across many decades and obey finite-size scaling (see Figure \ref{avalanches}c, Figure \ref{avalanches}d as well as Appendix G).  More specifically, one can fit exponent values $\tau \approx 2.1(1)$, $\alpha \approx 2.5(1)$ and $\gamma ^{-1} \approx 0.75(5)$, respectively \cite{alstott2014,clauset2009}. \new{Observe also that there is always a peak, corresponding to anomalously system-size spanning avalanches, but these ``bumps'' scale with system size in a scale invariant way (much as in self-organized bistability \cite{SOB})}. Analogous results are found for 2D lattices ---even if with slightly different exponent values: $\tau \approx 1.7(1)$, $\alpha\approx1.9(1)$ and $\gamma ^{-1} \approx 0.75(5)$, which are compatible with the values reported in experimental works \cite{Copelli-PRL}). In both cases, the scaling relation holds and the corresponding values of $\gamma$ coincide with its \new{seemingly robust value} reported in \cite{Copelli-PRL}. \new{Thus far we do not have a proper analytical understanding of these numerical values and their possible universality.}

As a complementary measure of complexity at the different types of transition points, we also computed the probability distribution of the inter-spike intervals (ISI), and its associated coefficient of variation (CV). Only around the HT transition, within the bistability region, we found broad ISI distributions with significant coefficients of variation $\cv > 1$ (see Appendix G). Thus, even if type-I and type-II synchronization transitions are well-known to exhibit signatures of criticality such as finite-size scaling (see e.g. \cite{Ohta-Sasa} and \cite{Kuramoto-FSS}, respectively) ---unlike to HT transitions--- they are not able to generate the higher levels of complexity required for scale-free avalanches and large dynamical repertoires as observed in the cortex.

\section{Discussion: on hybrid-type synchronization, criticality, and universality}

One could wonder whether the rich phenomenology emerging around the HT transitions stems from any of the three special codimension-2 bifurcation points that appear in the phase diagram (Figure \ref{phase_diagram}). Remarkably, the saddle-node-loop (SNL) bifurcation has been previously argued to be necessary for the generation of high variability and dynamical richness in neural networks \cite{SchreiberSNL,SchreiberHomoclinic}, and \new{it has also been established that the crucial features of dynamically-rich neural networks} stem from a phase diagram organized around a Bogdanov-Takens bifurcation \cite{Cowan2016,Benayoun}. As illustrated in the actual phase diagram (see Figure \ref{phase_diagram} and its 2D counterpart in Appendix H) the bistability region is rather small in the parameter space, so that all the discussed codimension-2 points are very close to each other. Thus, providing a clear-cut computational answer to the question above is a hard problem. Numerically, we can conclude that scale-free avalanches appear when entering the synchronous phase within the bistability region in the close vicinity of such codimension-2 bifurcations.

An aspect that needs further analysis is the relationship between criticality and bistability: these two features are usually opposed to each other, as they correspond to either continuous or discontinuous phase transitions, respectively  \new{(nevertheless, we should remind that scale-invariant avalanches can also appear at discontinuous phase transitions \cite{SOB, SOC+SOB}).}  However, the onset of synchronization in the HT transition occurs within a region of bistability. The intertwining between criticality and bistability \new{could well be} at the basis of the very rich dynamical repertoires in this case.  \new{It is very plausible, that the above-described LG theory, and possibly other models \cite{Zhou-2020}, exhibit scale-free avalanches at the edge of synchronization, since some kind of bistability is also present around the synchronization transition. }

Remarkably, the reported complex triangular-shaped structure \new{---with its bifurcation lines and co-dimension-2 points---} is rather universal and emerges in other models exhibiting a both type-I and type-II transitions (see e.g. \cite{ML-bifurcations,Theta, Pazo-PRX2015,Izhikevich-book}); in particular, it appears in the paradigmatic and broadly used (Wilson-Cowan) model of excitatory-inhibitory networks \cite{Borisyuk}.  Thus, the phenomenology discussed here is rather universal and not model-specific. It is noteworthy that other scenarios have been described to connect lines of type-I and type-II transitions \new{ --e.g., subcritical Hopf bifurcations followed by a fold of limit cycles--} which also involve a regime of bistability (see \cite{Copelli-2019}).  \new{Note that the phenomenology underlying these alternative scenarios is, at its core, similar to the one displayed at the HT synchronization transition, including collective excitability and some codimension-2 bifurcation points.}  The exciting possibility that scale-free avalanches can also emerge in such cases will be investigated elsewhere.

\newr{An aspect of our model that might need to be modified to better reproduce the results in the Landau-Ginzburg theory \cite{LG} is that here there are no true absorbing states, i.e. states from which the system cannot exit (neither as consequence of the deterministic dynamics nor as the result of stochasticity). In particular, the noise amplitude in \cite{LG} vanishes in the absence of activity, while in the present model such an ingredient cannot be easily implemented: large fluctuations can always excite individual nodes due to the absence of any bona-fide absorbing state.  On the other hand, it is well established that absorbing states are needed to generate branching-process exponents (see e.g. \cite{Serena-BP}). Thus, further work is still needed to elucidate what happens in modelling approaches as ours if absorbing states are included. This problem will be tackled elsewhere.}

Even if the minimal model studied here is exceedingly simple to be a realistic model of the cortex, it can provide us with insight on the basic dynamical mechanisms needed to generate  its complex dynamical features. Furthermore, it is well-established that diverse control, self-organization 
(or ``homeostatic'')
mechanisms are able to regulate a network to lie around \newr{some target regime or point \cite{JABO2,JABO1,Kinouchi,SOC+SOB}}. Thus, a cortical neural network with dynamics akin to that of the present simple model could be self-organized to the region near the HT synchronization transition \cite{SOB,PRR,SOC+SOB}, and by doing so, it could rapidly shift its behavior from synchronous to asynchronous, to collective excitability, or up-and-down transitions in a dynamical way, allowing for an extremely rich and flexible dynamical repertoire derived from operating at such an ``\emph{edge of the edge}''.  We hope this work opens the door for novel research lines, including renormalization group analyses \cite{Bialek-RG}), paving the way to the long-term goal of constructing a statistical-mechanics of the cortex.
 
\begin{acknowledgments}
  MAM acknowledges the Spanish Ministry and Agencia Estatal de investigaci{\'o}n (AEI) through grant FIS2017-84256-P (European Regional Development Fund), as well as the Consejería de Conocimiento, Investigación  Universidad, Junta de Andalucía and European Regional Development Fund, Ref. A-FQM-175-UGR18 and SOMM17/6105/UGR for financial support. V.B and R.B. acknowledge funding from the INFN BIOPHYS project.
  We also thank Cariparma for their support through the TEACH IN PARMA project. We thank S. di Santo, G. Barrios, D. Paz\'o  and J. Zierenberg for very valuable discussions and comments. 
 \end{acknowledgments}

\appendix

\section{Definition and measurement of avalanches}

\begin{figure}[hbtp] \begin{center}
  \includegraphics[width=1.0\columnwidth]{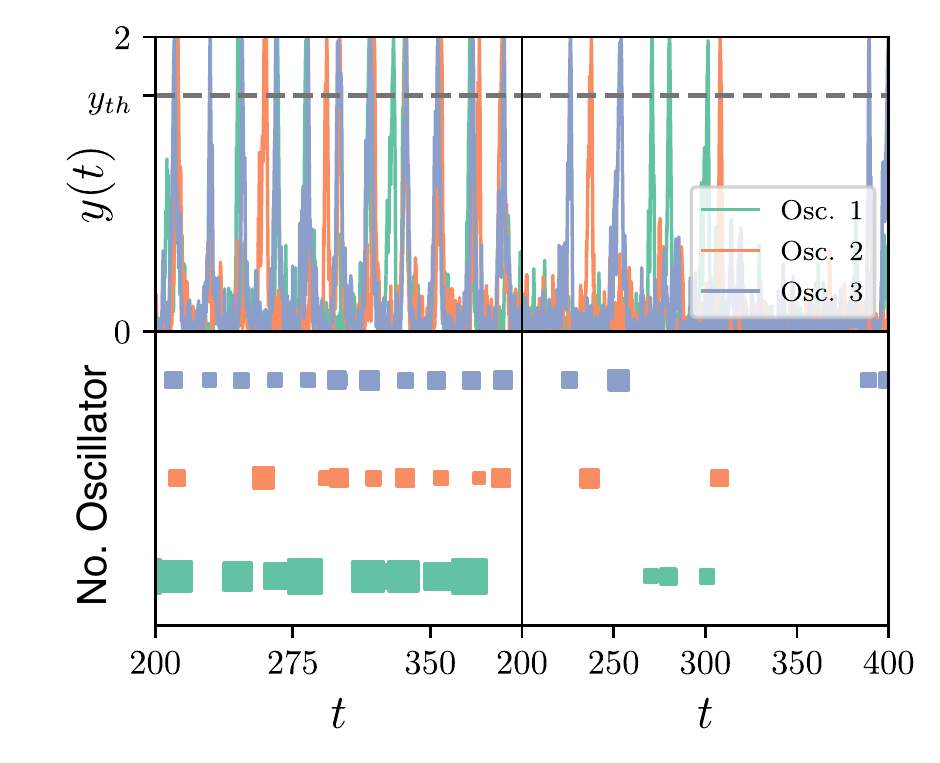}
  \caption{\new{\textbf{Construction of raster plots.} The individual noisy oscillators can display spike-like events; these can be detected by translating their phases into ``activities'' via the transformation $y(t)=1+\sin\varphi(t)$; then, by thresholding, the activity variable $y(t)$  one can define ``spikes'' as done, e.g., in local-field-potential measurements. The top panel shows the activity of $3$ randomly selected oscillators and the bottom one their corresponding spiking events as a function of time. The symbol size of each event accounts for the total time-integrated activity over the threshold. The left and right cases correspond to the synchronous phase and the bistability region, respectively.}}\label{spikesdefinition}
  \par\end{center}
\end{figure}

The protocol to measure neuronal avalanches is based on the one first proposed by Beggs and Plenz \cite{BP2003}, which  has been widely employed in analyzing both experimental and theoretical data (e.g. \cite{Schuster, Hahn2010, Copelli-PRL, LG, Benayoun}). This protocol allows one to study the structure of the spatiotemporal clusters of neuronal activity, e.g., spikes in individual neurons or peaks of the negative local field potentials.  Here, we discuss how the protocol is adapted for the case considered in this paper of phase oscillators, \new{illustrated in Fig.} \ref{spikesdefinition}.

\begin{enumerate}
\item For each unit $j$, one needs to define its ``activity''.  \new{Once the activity is defined}, a ``spike''  of a given oscillator $j$ is a transient event occurring whenever its phase $\varphi_j(t)$ crosses a given threshold value \new{which corresponds to a significant activity}. In particular, it is possible to define the activity as $y_j = 1+\sin\varphi_j$, that takes large (resp. small) values (close to $2$, resp $0$) when the phase crosses $\pi/2$ (resp. $-\pi/2$).

\item A threshold $y_{\text{th}}$ is defined such that whenever $y_j$ crossed, a ``spiking event'' starts to be tracked; it finishes when $y_j$ goes again below threshold. The total
  integral of $y_j - y_{\text{th}}$ along such a large-activity time window is the size $s_j$ of the local event at the initiation
  time $t_j^k$. $k=1, 2...$, etc., label the sequence of spikes.
  
\item The complete set of spikes for all units $j$ (at times $t_j ^k$ and sizes $s_j ^k$) defines a raster plot employed for the subsequent analyses.

\item The \new{inter-event interval (IEI)} between all couples of consecutive spikes (regardless of which unit generated them)  is computed, and its probability distribution $P(IEI)$ and its average value $\langle IEI \rangle$ are computed.  A time scale $\Delta t = \langle IEI \rangle$  is used to discretize the raster plots in time bins.
  
\item An avalanche is defined as a series of spikes in between two empty bins (with no spike) such that all consecutive time bins include  some activity. The sum of all $s_j ^k$ in between such two empty bins is the avalanche size $S$, while avalanche duration $T$ is defined as the time elapsed between the two limiting empty bins.

\item The probability distribution function (histogram) of avalanche sizes and durations is then computed.  \end{enumerate}

The empirical detection of avalanches in noisy data and/or in continuous time series is often exposed to some potential pitfalls that are important to underline.
\begin{itemize}
  \item
First of all, it relies on an arbitrary choice of a threshold for detecting ``spikes'' of activity.
In our case, selecting a threshold $y_{\text{th}}$ (e.g. $y_{\text{th}} = 1.6$) ensures that background noise around the fixed point is filtered and only significant excursions in the phase value are considered.

\item
  It also relies on the choice of the time bin size as the average inter-event interval; this choice has been made in accordance with the usual one in neuroscience analyses \cite{BP2003, Schuster, Yu, Copelli-PRL}.
  
\item
Let us also remark that, instead of using the activity $y_j$ with its corresponding threshold, we could have \new{alternatively} defined a ``spike'' each time oscillator hits a predefined phase value, such as $\pi/2$. \new{This method does not allow to integrate the ``size'' for each event, but it makes no difference with the previous one. However, in our particular case, we seek to mimic the dynamics by \cite{LG} and actual LFP recordings, where each event has a size. Therefore, we stick to the threshold $y_{\text{th}} = 1.6$ for simulations}. 


\item Let us finally stress that computational model analyses, like the ones reported here, do not suffer from \emph{severe} subsampling effects that may strongly impair empirical measurements of avalanches \cite{Subsaampling1,Subsaampling2, Subsampling3}. \end{itemize}

\section{Avalanches in the annealed Kuramoto model}

\begin{figure}[hbtp]
\begin{center}
   \includegraphics[width=1.0\columnwidth]{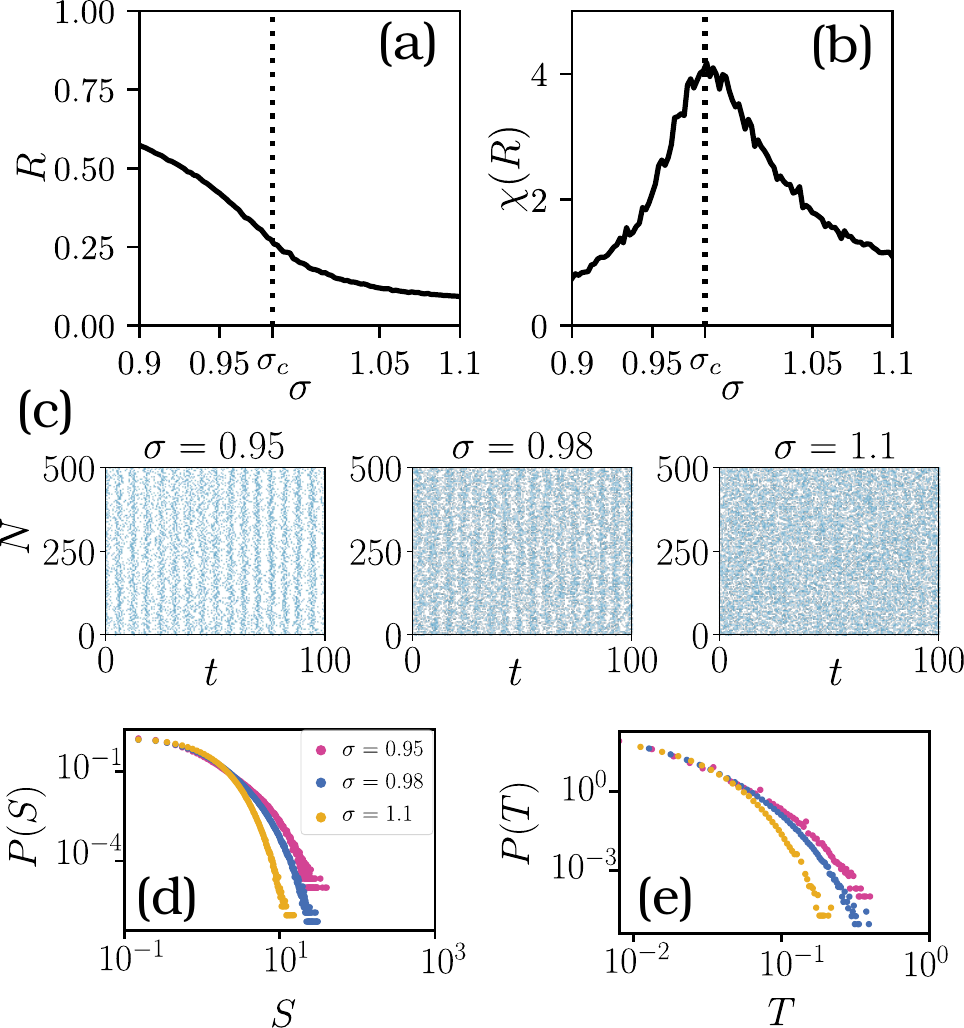}
  \caption{\textbf{Avalanche statistics in the annealed Kuramoto model on fully connected networks.} (a) Kuramoto order parameter ($R$) as a function of the noise intensity $\sigma$ for a finite network size $N=500$. (b) The Kuramoto critical point is defined as the point of maximum variance of the order parameter, $\chi$, which occurs at $\sigma_c = 0.98$ (vertical dashed line). 
(c) Raster plots of the annealed Kuramoto model at the synchronous phase ($\sigma=0.95$, left plot), the critical point ($\sigma_c=0.98$, central plot) and the asynchronous phase ($\sigma=1.1$, right plot). (d) Distributions of size events in the three phases for the same representative values of $\sigma$ as above. (e) Distribution of time events in all the three phases for the same parameter values. Let us underline the lack of power-law distributions at criticality, i.e. when the system undergoes a collective Hopf bifurcation. Parameter values: $\omega=1$ and $J=1$.}\label{AvK}
  \end{center}
\end{figure}

Without loss of generality, we fix $J=1$ and $\omega=1$ ---that is not relevant as a change of variables to a co-rotating reference frame can be used to set $\omega=0$--- leaving $\sigma$ as the only free parameter. Then, $\sigma_c=1$ indicates the critical point for infinitely large systems. Due to the finite size $N$ of the considered networks, the precise location of the critical point needs to be computationally estimated; in particular, as usually done in finite-size scaling analyses, $\sigma_c$ is estimated as the value of $\sigma$ such that the variance of the order parameter $R$ is maximal \cite{Acebron,LG}. As shown in Figure \ref{AvK}a and Figure \ref{AvK}b, showing results of our computer simulations for a system of $N=500$ units, the critical point is located at $\sigma_c \approx 0.98$ (and shifts progressively towards $1$ as $N$ is increased). Observe that at the estimated critical point, owing to finite-size effects, the level of synchronization is $R\simeq0.2$. This is illustrated in Figure \ref{AvK}c, which shows three characteristic raster plots within the synchronous phase $\sigma=0.9$, the critical point $\sigma_c(N=500) \approx 0.98$ and the asynchronous phase $\sigma=1.1$, respectively.

As an important technical remark, we should emphasize that in numerical simulations of the annealed
Kuramoto model, for any finite size $N$, the integration step $\delta t$ needs to be small enough as to have sufficient time resolution, i.e.  $\delta t < \Delta t = \langle \text{IEI} \rangle$, so that avalanches can be measured. Observe that this might depend on $N$; in particular, in the asynchronous phase, as the number of neurons $N$ grows, the $\langle \text{IEI} \rangle$ decreases, and thus one needs progressively smaller integration steps to measure avalanches. In the limit $N\rightarrow \infty$, the asynchronous raster plot has an average \new{interevent interval $\langle \text{IEI}\rangle \rightarrow 0$ and}, consequently, one could say that avalanches are not well defined if a fixed time bin was considered.  On the other hand by considering sufficiently small $\delta t$'s for each case,  our computational analyses --summarized in Figure  \ref{AvK} panels (d) and (e)-- show that the avalanche statistics do not exhibit heavy tails. 

\new{Finally, a careful mathematical analysis of this model and its avalanches in the thermodynamics limit will be addressed elsewhere.}

\section{Bifurcations in the single unit of the Landau-Ginzburg model}
The Landau-Ginzburg model \cite{LG} was pioneer in proposing that scale-free avalanches occur at the edge of a synchronization phase transition. It relies on a model for the dynamics of individual mesoscopic regions in the cortex. Each such region (or ``unit'') is characterized by two dynamical variables: its level of (mesoscopic) neuronal activity $\rho(t)$ and the amount of available synaptic resources $R(t)$. The dynamics (at a deterministic level, i.e. excluding fluctuations) is described by \begin{equation} \begin{cases}
  \dot{\rho}(t)=(R(t)-a)\rho+b\rho^2 -\rho^{3}+h\\
  \dot{R}(t)=\frac{1}{\tau_R}(\xi-R)- \frac{1}{\tau_D} R\rho \end{cases} \label{PNAS} \end{equation} where $a,b>0$ are constants, $h$ is a very small external driving (that can be set to $0$ in the absence of external stimuli), $\eta(t)$ is a zero-mean Gaussian white noise, and $\xi$ is the maximum amount of available synaptic resources, which serves as a control parameter which regulates the system state. In the second equation $\tau_R$ and $\tau_D$ represent the time scales of recovery and depletion of synaptic resources, respectively.

  \begin{figure}[hbtp] \begin{center}
  \includegraphics[width=1.0\columnwidth]{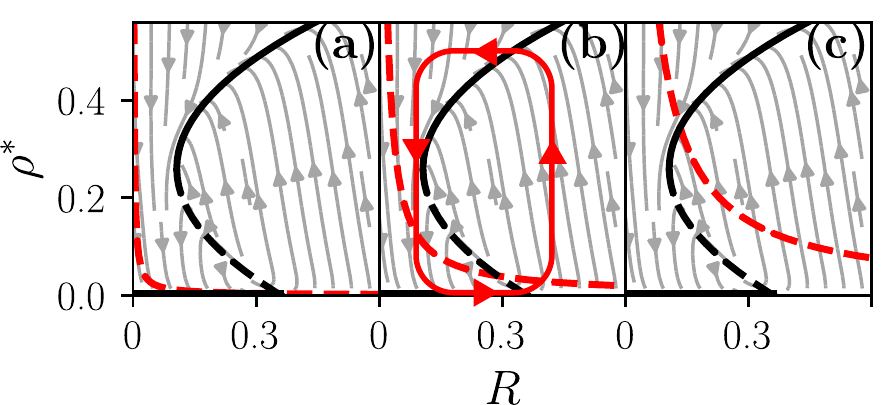}
  \caption{\textbf{Nullclines of the single mesoscopic unit of the Landau-Ginzburg model \ref{PNAS}.}
    Characteristic  flow diagrams and nullclines of the Landau-Ginzburg mesoscopic unit, for three different values of $\xi$. The nullclines for the activity $\rho$  (black lines) can display two types of solutions depending on the value of the available resources, $R$: up ($\rho \neq 0$ for large values of $R$), down ($\rho = 0$ for small values of $R$) and bistability between these two for intermediate values of $R$. Black dashed lines represent unstable fixed points $\rho^*$. Nullclines $\dot R = 0$ are represented by red dashed lines. The only stable fixed point for low values of the order parameter, $\xi$, is the absorbing state $\rho^* = 0$ (black point). (b) When $\xi$ is increased, the $R$-nullcline intersects the unstable branch of the $\rho$-nullcline, giving rise to a limit cycle (red solid line). (c) When $\xi$ is large enough, the up-state fixed point becomes the only stable solution.}\label{PNAS_nullclines}
\end{center}
\end{figure}

Observe that the equation for the activity $\rho$ --assuming the amount of synaptic resources $R$ is fixed-- is the minimal form of a first-order phase transition with hysteresis (or saddle-node bifurcation).  It displays a quiescent (or ``down'') state $\rho = 0$ when $R \leq a$, and an active or ``up'' state for $R > a$. On the other hand, the second equation accounts for the dynamics of the level of synaptic resources and includes a slow charge/recovery term (dominating when activity is low), and a fast discharge/consumption, which dominates the dynamics in the presence of activity, $\rho \neq 0$.

\begin{figure}[hbtp]
\begin{center}
  \includegraphics[width=1.0\columnwidth]{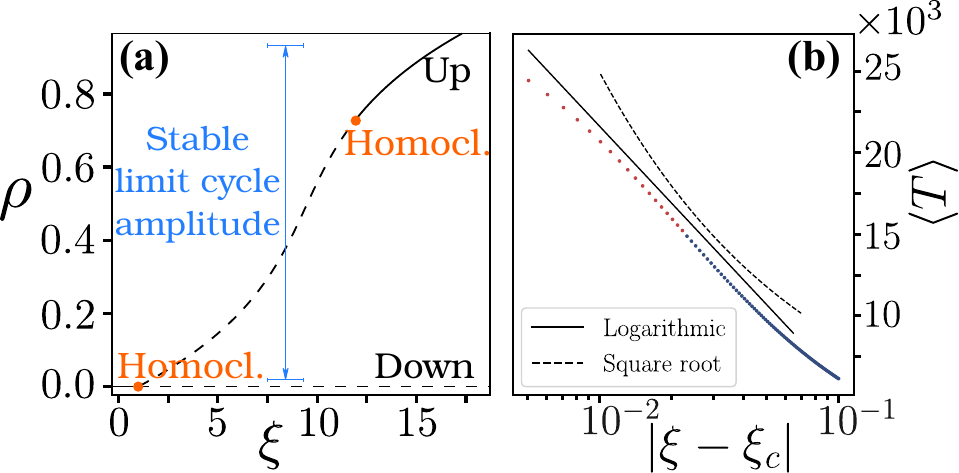}
  \caption{\textbf{Bifurcations of the single unit dynamics of the Landau-Ginzburg model.}  Stable fixed points of Eq.(\ref{PNAS}) are represented as a continuous line, while unstable ones correspond to dashed lines. Parameters: $a=1.0$, $b=1.5$, $\tau_R=10^3$, $\tau_D=10^2$, $h=0$. (a) As the control parameter $\xi$ is increased, the down-state fixed point loses its stability via a homoclinic bifurcation. (b) Period of oscillations as a function of the control parameter $\xi$. The set of equations \ref{PNAS} was integrated for a fixed long time to compute the period as the average time between spikes (jumps to the up branch). Points marked in red indicate that the number of spikes used for computing the average period were relatively low (because of costly statistics).  Note that as $\xi \rightarrow \xi_c$, larger simulations are required (increasing integration error). The data are much better fitted by a logarithm (characteristic of homoclinic bifurcations) than by a square-root fit (characteristic of SNIC bifurcations) \cite{Strogatz-book}. All numerical solutions are found using Wolfram Mathematica.}\label{PNAS_bif}
\end{center}
\end{figure}

A simple analysis of Eqs.(\ref{PNAS}) shows that the system behavior depends on the value of the maximum allowed synaptic resources, $\xi$ (see Figure  \ref{PNAS_bif}). If $\xi < a$, the only fixed point is a quiescent state of low activity.  For larger values of $\xi$, the two nullclines of Eq.(\ref{PNAS}) intersect at an unstable fixed point, giving rise to a limit cycle, i.e. to \emph{relaxation oscillations} in which both $\rho(t)$ and $R(t)$ oscillate. Finally, if $\xi$ is large enough, an up state (fixed point with non-vanishing activity) emerges. For more details we refer to \cite{LG} and \cite{PRR}.

Figure \ref{PNAS_bif} illustrates the bifurcation diagram of Eq.(\ref{PNAS}) as the control parameter $\xi$ is varied. In agreement with what just described, values $\xi<a$ leads to a ``down''  steady state with vanishing activity and non-depleted synaptic resources, i.e. $R=\xi$.  At $\xi_c = a$ there is an \emph{ infinite-period homoclinic  bifurcation} into a limit cycle.   To determine if this bifurcation is homoclinic or rather a saddle-node into an invariant circle (SNIC) one, we have explicitly measured the average period between oscillations and plotted against the control parameter. The result displays a logarithmic decay of the period (see  Figure  \ref{PNAS_bif}b), a typical feature of homoclinic bifurcations \cite{Strogatz-book}. Finally, as the control parameter is further increased, one encounters another homoclinic  bifurcation at which the limit cycle disappears, giving rise to an ``up'' fixed point.

\section{A note on different types of excitability and  bifurcations}
A system is defined as \emph{excitable} when it presents a single, stable equilibrium, but a sufficiently strong input can drive the system in a large excursion in the phase space before returning later to the stable fixed point \cite{Izhikevich-book, Ojalvo-review, Jaeger-excitability}. Many physical and biological systems exhibit excitability \cite{Ojalvo-review}. Excitability is a concept of particular importance in the context of neuroscience, where neurons are at rest, and a super-threshold signal is able to evoke a significant response (e.g., a spike), returning at the end to the resting state.  Different types of neurons may respond differently to the same input, leading to the so-called ``excitability classes'' \cite{Izhikevich-book, Jaeger-excitability}.
 \begin{figure}[hbtp] \begin{center}
  \includegraphics[width=1.0\columnwidth]{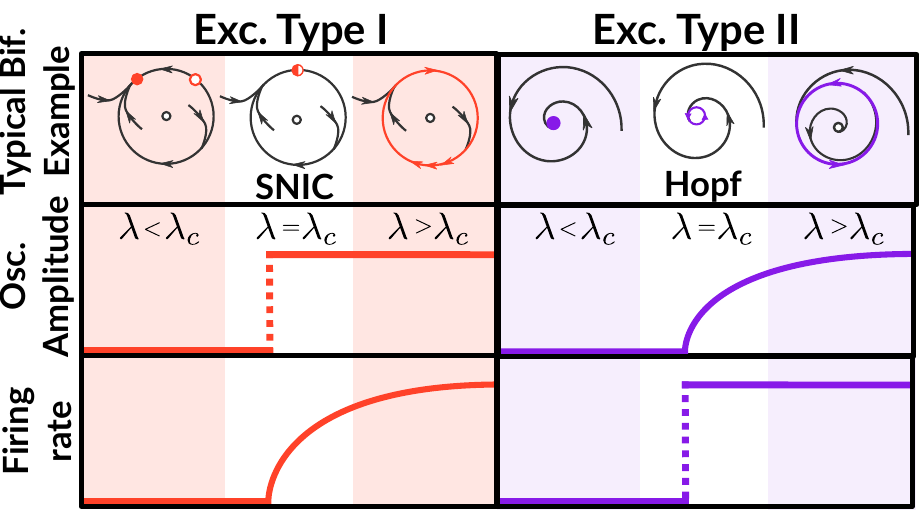}
  \caption{\textbf{Excitability and bifurcations.} Side-by-side comparative sketch of infinite-period (SNIC) and Hopf bifurcations, which are representative examples of type-I and type-II excitability, respectively. Note that the firing rate is directly related with frequency of oscillations: vanishingly-small firing rates correspond to arbitrarily large firing frequencies. }
  \label{snichopf}
  \par\end{center}
\end{figure}
The most usual classes are excitability classes I and II \cite{Izhikevich-book}.
\emph{Type-I excitability} is characterized by continuous growth of
the spiking rate when the input current is continuously increased, while \emph{Type-II excitability} involves a sudden jump in the spiking rate under the same circumstances.
In bifurcation theory, type-I excitability corresponds to situations  where the corresponding limit cycle appears with a vanishing frequency, i.e. infinite-period bifurcations, while in type-II excitability, limit cycles emerge with a finite, non-vanishing frequency \cite{Izhikevich-book}.

Two representative examples of bifurcations corresponding to classes I and II are the SNIC and Hopf bifurcation, respectively, as sketched in Figure \ref{snichopf}. The homoclinic bifurcation is common in neuronal models \cite{SchreiberHomoclinic} and belongs to class I neuronal excitability, as the SNIC bifurcation. Both are differentiated only by the critical exponents of the firing rate. A comprehensive summary of the relationship between excitability classes and bifurcations can be found, for example, in \cite{autapses}.

It should be noted that the mean-field phase diagram described in the main text includes these two main types of excitability near the bistability region, where the system behaves as collectively excitable.  The bistability region is surrounded by other types of bifurcations such as saddle nodes and \emph{codimension-2 bifurcations}, where lines of standard (codimension-1) bifurcations intersect. Codimension-2 points, which in our case include Bogdanov-Takens, saddle-node-loop, and cusp bifurcations, can display richer dynamics \cite{Izhikevich-book, Cowan2016, SchreiberSNL, SchreiberHomoclinic}).

We have classified synchronization transitions using a nomenclature that resembles that of excitability classes. In type-I synchronization, oscillations emerge at the
transition point with zero frequency (infinite period) and finite amplitude, while in type-II synchronization, oscillations are born with a fixed non-vanishing frequency. The phenomenology becomes richer in "hybrid-type" synchronization transition, where the co-dimension 2 bifurcations and bistability are present.

\section{Mathematical analysis of globally coupled oscillators} \label{sec-order-par-eqs}

To assess the collective behavior of the system of coupled oscillators,
here we reproduce with detail the derivation of the equations for the order parameter and discuss different closure methods. Most of these calculations can be found in the literature, but we reproduce them here in a self-consistent way for the sake of clarity. Our model is described by the set of stochastic equations,

\begin{equation} 
\dot{\varphi_j}=\omega + a\sin\varphi_j + \dfrac{J}{M_j}{\sum_{i \in n.n.j} ^{M{j}}} \sin\left( \varphi_i - \varphi_j \right) + \sigma \eta_j(t). \label{SNIC-coupled} 
\end{equation}

It is convenient to remove one parameter, fixing e.g. $\omega=1$. Here, we also set $J=1$, leaving $a$ and $\sigma$ as the only free parameters. For completeness, we verify a posteriori that results are robust to changes in $J$ (Appendix F).

\subsection{Order-parameter equations} 

Equation (\ref{SNIC-coupled}) is very similar to the previously discussed annealed Kuramoto model, except for the additional term $a\sin(\varphi)$ which induces an inhomogeneity in angular velocity across the unit circle of each oscillator. As discussed in Appendix G (in particular, in Section \ref{sec-phase-diagram}), such an inhomogeneity makes the Kuramoto parameter inadequate to characterize the phase diagram of the present model as, for $a \neq 0$, there is a particular phase value around which each oscillator tends to spend most of the time. For uncoupled oscillators, this occurs for $\varphi=\pm\arcsin(\omega/a)$ where the angular velocity is minimal; this heterogeneity
in angular velocity leads to a non-vanishing value of the Kuramoto order parameter, even when oscillators are uncoupled or, more in general, when they are asynchronous.

In order to circumvent this problem analytically, it is possible to consider the hierarchy of  higher-order moments of the variable $e^{i\varphi}$, i.e., the so-called Kuramoto-Daido parameters, $Z_k$:
\begin{equation}
Z_k = \langle e^{ik\varphi} \rangle \equiv \frac{1}{N} \sum_{j=0} ^N  {e^{ik\varphi_j}} \equiv R_k e^{i\psi_k}
\end{equation}
where $k=1,2,...\infty$ and of which the Kuramoto order parameter $Z_1$ is a particular case.  For convenience, we will use either the notation in terms of amplitudes and phases ($R_k$ and $\psi_k$) to represent the complex-valued Kuramoto-Daido parameters $Z_k$.  Using standard trigonometric relations, Eq.(\ref{SNIC-coupled}) can be rewritten as a function of $Z_1(t)=R_1(t) e^{i \psi_1(t)}$, leading to the following set of Langevin equations,
\begin{equation} 
\dot{\varphi}_{j}(t) = \omega +a\sin\varphi(t) + JR_1(t)\sin\left(\psi_1(t)-\varphi_{j}(t)\right) + \sigma\eta\left(t\right) 
\end{equation}
where the mean-field nature of the coupling is evident.

In order to solve these equations, we employ a standard procedure to deal with coupled oscillators \cite{SK-1, SK-2, Pikovsky-book, Geier-gaussian, Pikovsky-reduction}. The first step is to consider a large number of oscillators, $N \rightarrow +\infty$, so that the system can be described in the continuum limit using the probability density to find an oscillator around any given phase value $\varphi$, i.e. $P(\varphi) d\varphi$. The following Fokker-Planck equation gives the evolution of such a probability density,

\begin{multline}
\partial_{t}P\left(\varphi,t \right) = \frac{\sigma^{2}}{2}\partial_{\varphi}^{2}  P\left(\varphi,t\right) - \\
-J\partial_{\varphi}\left[\left(\omega + a\sin\varphi +
  \frac{Z_1 e^{-i\varphi} + c.c.}{2i}  \right) P\left(\varphi,t\right)\right]  
\label{SNIC_FP}
\end{multline}

where the identity   $R_1\sin\left(\psi_1-\varphi\right) = (Z_1 e^{-i\varphi} + \bar Z_1 e^{i\varphi})/(2i)$ has been used to simplify the forthcoming algebra. As the density $P(\varphi,t)$ is periodic in the angle variable, it can be expanded in Fourier series:

\begin{equation}
  P(\varphi,t) = \dfrac{1}{2\pi} \sum_{k=-\infty} ^{+\infty} p_k(t) e^{i k \varphi},
  \label{Fseries}
\end{equation}
and $p_k = \bar{p}_{-k}$, where the bar stands for complex conjugate.
It turns out that the Kuramoto-Daido parameters coincide with these coefficients:
\begin{equation}
  Z_k = \int _0 ^{2\pi} P(\varphi,t) e^{ik\varphi} d\varphi = p_{-k}.
\label{zkpk}
\end{equation}

Plugging the series expansion (\ref{Fseries}) into the Fokker-Planck equation one obtains an infinite set of differential equations, one for each of the parameters $Z_k$, i.e. for the Kuramoto-Daido parameters $Z_k$ (observe that, as $Z_{-k} = \bar{Z}_k$, it suffices to analyze the order parameters with $k \geq 0$). To obtain differential equations for each $Z_k$, note that after performing the derivatives and doing some algebra, all the terms can be written as $(2\pi)^{-1} \sum_k f(Z_k, Z_{k+1}, Z_{k-1}, \ldots) e^{ik\varphi}$ for some function $f$. Then, since the exponentials $e^{ik\varphi}$ are the Fourier-basis elements, we can identify all parameters mode by mode, leading to an equation for the evolution of each Kuramoto-Daido order parameter \cite{Geier-gaussian, Pikovsky-cumulants}.
The resulting set of equations,  

\begin{multline}
\dot Z_k = Z_k(i \omega k - \frac{k^2 \sigma^2}{2} ) + \frac{a k}{2} \left( Z_{k+1} - Z_{k-1} \right) + \\ \frac{J k}{2} \left( Z_1 Z_{k-1} - \bar Z_1 Z_{k+1} \right),
\label{zk_system}
\end{multline}

constitutes an exact description of the system.

As reported below, we solved Eq.(\ref{zk_system}) including up to $k=50$ harmonics (i.e. imposing $Z_{51} = 0$) and monitored the order parameters (see Section \ref{sec-phase-diagram}).  We found an excellent agreement with direct simulations, including a small bistable phase \cite{SK-2, Geier-gaussian}, as shown in Figure  2C of the main text. {However, even if the bistable phase exists in the thermodynamic limit, its exact location is affected by finite-size effects in direct simulations.

In order to proceed analytically, 	given that all equations are coupled, one needs
to find a dimensional sound reduction or ``closure'' to truncate the infinite hierarchy \cite{Pikovsky-reduction, Makarem-Closure2017}. Different closures have been proposed in the literature of coupled oscillators. In what follows, we discuss three of them.

\subsection{Approximate solutions or closures}

In deterministic, noise-free systems, an exact solution to equation (\ref{zk_system}) is provided by the Ott-Antonsen ansatz \cite{OA}, which consists in writing $Z_k(t) = [Z(t)]^k$,
so that the first moment already contains all the relevant information. On the other hand,
the situation is more complicated for stochastic systems for which the Ott-Antonsen ansatz does not provide an exact solution \cite{Pikovsky-reduction, Pikovsky-cumulants}. In particular, using the Ott-Antonsen ansatz and writing $Z(t) = R(t)e^{i\psi(t)}$ leads to the system of equations

\begin{eqnarray}
\dot{R} &=&	\frac{1}{2}R\left[ J\left(1-R^{2}\right)-\sigma^{2}\right]-\frac{1}{2}a\left(1-R^{2}\right)\cos\psi, \nonumber \\
\dot{\psi} &=&	\omega+\frac{a\left(1+R^{2}\right)\sin\psi}{2R}.
\label{OA}
\end{eqnarray}

Remarkably, these equations are the same as those obtained by Childs and Strogatz in the case of deterministic oscillators with heterogeneous (quenched) frequencies distributed as a Lorentzian \cite{Strogatz-childs}. However, only the deterministic system is exactly solved by the Ott-Antonsen ansatz, while in the stochastic case it gives an approximation of order $\mathcal{O}(\sigma^2)$ \cite{Pikovsky-cumulants}.

In order to increase the precision of the closure, one can assume that the global phase is distributed as a (wrapped) Gaussian with some mean $\psi(t)$ and variance $\Delta(t)$ \cite{Geier-gaussian}: 

\begin{equation}
P(\varphi, t) = \frac{1}{\sqrt{2\pi\Delta}} \sum_{k=-\infty} ^{+\infty} \exp\left[ -\frac{(\psi-\varphi+2\pi k)^2}{2\Delta} \label{wrapped-gaussian} \right]
\end{equation}

The Kuramoto-Daido parameters can be explicitly computed via direct integration,

\begin{equation}
Z_k(t) = \int d\varphi P(\varphi, t) e^{ik\varphi} = e^{-\frac{1}{2}k^2\Delta(t)} e^{ik\psi (t)}.
\label{zk-gaussian}
\end{equation}
Plugging this ansatz into Eq.(\ref{zk_system}) leads to 
\begin{eqnarray}
\dot \psi(t) &=& \omega + a e^{-\Delta/2} \cosh\Delta \sin\psi, \\
\dot \Delta(t) &=&  \sigma^2 + 2\sinh\Delta \left( a e^{-\Delta/2}\cos\psi - J e^{-\Delta}  \right).
\label{gaussian}
\end{eqnarray}
Observe that Eq.(\ref{zk-gaussian}) allows one to write $Z_k$ as a function of only the first mode \cite{Pikovsky-cumulants}, $Z$, giving a functional form very similar to the Ott-Antonsen ansatz:
\begin{equation}
Z_k = |Z|^{k^2-k} Z ^k 
\end{equation}
Let us remark that the Ott-Antonsen ansatz is equivalent to the assumption of a Lorentzian distribution for the angles. That is, changing Eq.(\ref{wrapped-gaussian}) to a Lorentzian distribution and following the same procedure, one recovers the Ott-Antonsen ansatz.  The wrapped Gaussian approximation is slightly superior to the Ott-Antonsen one, but, as we will see shortly, it is not good enough as to generate a precise phase diagram (see Figure \ref{closures-vs-sim} below for a comparison).

A way to go beyond the Ott-Antonsen and the Gaussian ansatzes is to consider more harmonics in the expansion. Tyulkina et al \cite{Pikovsky-reduction, Pikovsky-cumulants} proposed to use the circular cumulants of the order parameters to generate a better closure. The advantage of the cumulant expansion is that all cumulants, except $\chi_1 = Z$, vanish when the Ott-Antonsen ansatz is selected, so choosing additional non-zero cumulants gives rise to systematic corrections to the Ott-Antonsen solution, order by order in such an expansion. In particular,  the first three cumulants are given by $\chi_1 = Z$, $\chi_2 = Z_2 - Z ^2$, $\chi_3 = (Z_3 - 3 Z Z_2 + 2Z ^3)/2$. Selecting $Z$ and $\chi_2$ to be different from $0$ but fixing $\chi_3=0$ gives $Z_3 = Z^3 + 3Z\chi_2$, effectively closing the infinite system.\footnote{In stochastic processes, the closure \unexpanded{$\langle x^3\rangle \simeq \langle x \rangle ^3 + 3\langle x\rangle \langle x^2 \rangle$} is also called \emph{Gaussian closure} because it assumes that the variable $x$ follows Gaussian statistics. Note that in the case under study, assuming that $Z$ obeys Gaussian statistics and considering that $\psi$ is Gaussian distributed are two different
approximations.}
Substituting this last in the equation for the first three harmonics, the resulting system reads:
\begin{align}
\dot Z =& \frac{1}{2} \left( J - \sigma^2 + 2i\omega \right) Z + a\left( Z^2 -1 + \chi_2  \right) \nonumber \\ 
 & - J \left(Z |Z|^2 + \chi_2 \bar Z \right), \nonumber \\
\dot \chi_2  =& 2\chi_2(i \omega + a Z) - \sigma^2(2\chi_2+Z^2) - 2 J \chi_2 |Z|^2. 
\label{tyulkina}
\end{align}
Note that imposing $\chi_2=0$ leads to the Ott-Antonsen ansatz, as expected. This closure provides us with a noticeable quantitive improvement on where the Hopf and SNIC bifurcations are located, which coincide very well with the results of numerical integrations. \new{However, simple parameter inspection did not render any bistability using the reduction by Tyulkina \emph{et al.} Given the size of the bistable region in phase space, finding the bistability probably needs a more systematic study -such as the one we did solving the system for $k=50$ harmonics, where bistability is clear as we showed in Fig. \ref{phase_diagram}. Although} losing the possibility of bistability in analytical approaches relying on some closures is a well-known problem in stochastic processes \cite{Makarem-Closure2017}.

\section{Bifurcation analysis of the Ott-Antonsen equations} 

To gain analytical insight into the structure and topological organization of the phase diagram, here we analyze the bifurcation diagram of the approximation provided by the simpler Ott-Antonsen closure, eqs. (\ref{OA}), i.e.: 
\begin{eqnarray}                                                                                                                       \dot{R} &=& \frac{1}{2}R\left[J\left(1-R^{2}\right)-\sigma^{2}\right]-\frac{1}{2}a\left(1-R^{2}\right)\cos\psi, \nonumber \\ \dot{\psi} &=& \omega+\frac{a\left(1+R^{2}\right)\sin\psi}{2R}.
\end{eqnarray}                               
                                                                    
Observe first that, for a fixed value of $R$, the equation of the collective phase $\psi$ is the normal form of a saddle-node into an invariant circle (SNIC) bifurcation. On the other hand, the equation for $\dot R$ is almost the same as in the annealed Kuramoto model, but adding a perturbation proportional to the ``excitability parameter'' $a$. The above set of equations is difficult to study analytically, but its bifurcations can be obtained following the same procedure of Childs and Strogatz who studied this system with a fixed value of $\sigma = \sqrt{2}$ \cite{Strogatz-childs}.

The main idea is as follows: in the annealed Kuramoto limit ($a=0$) the system undergoes a Hopf bifurcation; on the other hand,  individual (uncoupled) oscillators, i.e., for $J=0$, exhibit a SNIC bifurcation. Hence, by continuity in solutions, we expect two branches of these two types of bifurcations to be present in Eq.(\ref{OA}).

Calling $Q$ the Jacobian at a fixed point, at Hopf bifurcations $\tr Q = 0$ while in a saddle-node $\det Q=0$ \cite{Strogatz-book}. Thus, imposing one of these conditions, together with the fixed point equations $\dot R = \dot \psi = 0$, leads to a set of equations for the parameters of the system as a function of the fixed point values, $R^*$ and $\psi^*$.
Since such values are bounded, one can use these equations as parametric equations of the bifurcation curve, without computing explicitly the values of the fixed points and their stability.

Let us start with the Hopf bifurcation. Between parameters and fixed points, there are $6$ unknowns: $R^*$, $\psi^*$, $\omega$, $a$, $J$ and $\sigma$. Remember that in the simulations we fixed $\omega=J=1$ to leave $a$ and $\sigma$ as the only free parameters. In what follows, we obtain equations for the bifurcations, written in a parametric form $a=a(R^*, \psi^*, \omega, J)$ and $\sigma=\sigma(R^*, \psi^*, \omega, J)$. After some algebra, it turns out that not all the dependences are necessary.  
Solving for $\dot R = \dot \psi = \tr Q = 0$, there are $3$ remaining parameters. We can choose any parameters to solve for, but it turns out (as shown in \cite{Strogatz-childs}) that solving for $R$, $\cos\psi$ and $\sin\psi$ is highly convenient. Note that these are only two parameters, since the sine and cosine are not independent function as $\sin^2\psi+\cos^2\psi=1$.
Of course, one could have tried to directly solve for $\psi$ and $a$, but it is easier to obtain expressions for the trigonometric functions and then extract $a$:
\begin{equation}
a_H = \sqrt{\frac{J-\sigma^2}{J+\sigma^2} }  \frac{\sqrt{4\omega^2(J+\sigma^2)^2+J^2(J-\sigma^2)^2}}{2J}
\label{oa-hopf}
\end{equation}
Note that, in this case, we obtained a parametric curve for the Hopf bifurcation, $a_H=a_H(\omega, J,\sigma)$, without requiring specific knowledge about the location of the fixed points. In particular, as $J$ is kept fixed,  it is possible to derive  a curve $a_H=a_H(\omega, \sigma)$.

In what respects saddle-node bifurcations, the calculation is a bit more involved since solving for the same $3$ variables gives high-degree polynomials for $R$ that cannot be explicitly solved. For this reason, we choose to solve for $\omega$, $\cos\psi$ and $\sin\psi$, since in this way the parameters do not depend on $\psi$. After solving and simplifying the resulting equations, one obtains:

\begin{eqnarray}
  \omega_S &=& \frac{(1+R^2)^{3/2}}{2(1-R^2)^2} \cdot  \\
  & & \sqrt{J(1-R^2)\left( 2\sigma^2 - J(1-R^2)^2 \right) - \sigma^4 (1+R^2) },
  \nonumber \\
   \label{oa-saddle}
a_S &=&	\frac{\sqrt{2}R^2}{(1-R^2)^2} \cdot \\
& & \sqrt{\left(J(1-R^2)-\sigma^2 \right) \left(2\sigma^2-J(1-R^2)^2 \right) \nonumber}.
\end{eqnarray}

Since $0\leq R\leq 1
$, and one needs to explore all the possible fixed points, there are two free parameters to choose. Given that $J$ is kept fixed in our simulations, we dismiss it and obtain $\omega$ and $a$ as functions of $\sigma$. \new{The saddle-node lines enclose a ``collective excitable'' phase, which is type-I excitable. This can be easily seen by realizing, as discussed above, that phase dynamics are again of the form $\dot \psi = \omega + c \sin\psi$ for fixed Kuramoto order parameter $R$, with the angular speed taking the role of the external input usually employed to classify excitability classes}.

\begin{figure}[hbtp]
\begin{center}
  \includegraphics[width=1.0\columnwidth]{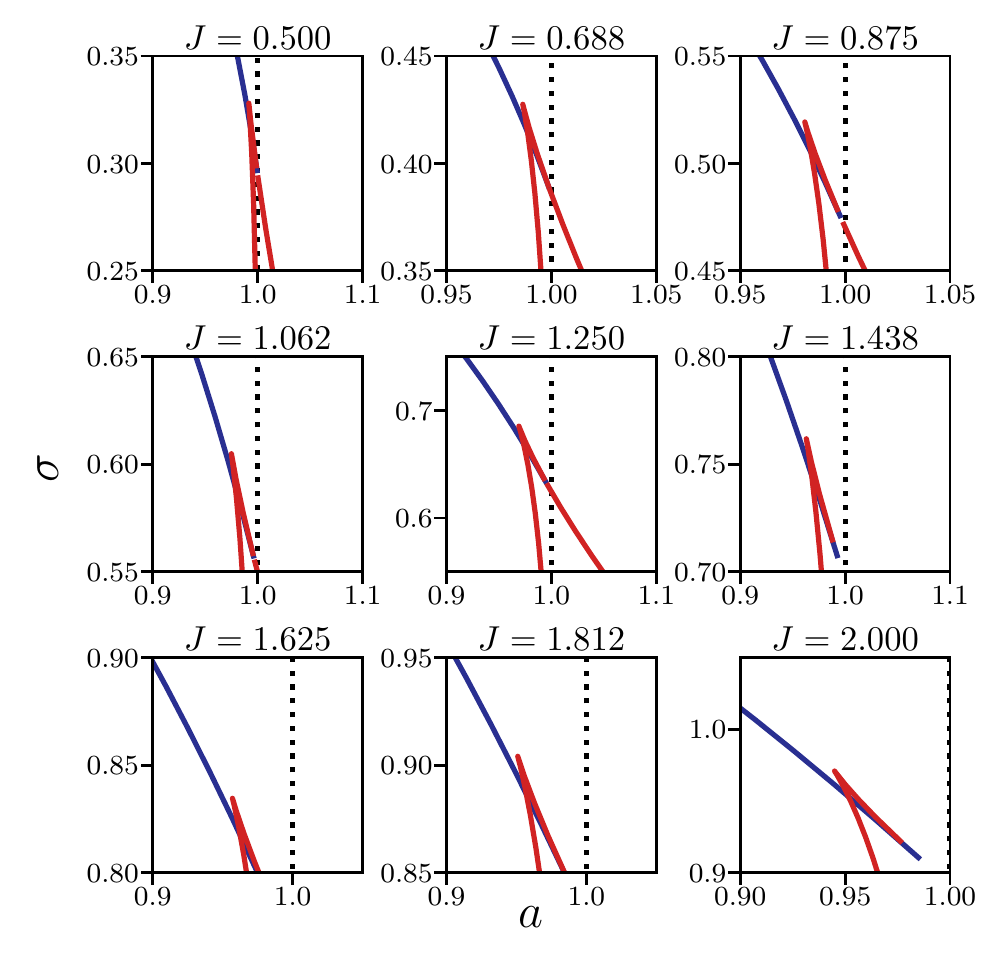}
  \caption{\textbf{Bifurcations in the Ott-Antonsen approximation.} Representation of the solutions of the equations (\ref{oa-hopf}) and (\ref{oa-saddle}) for different values of the coupling constant $J$.  Blue lines describe branches of Hopf bifurcations, while the red lines correspond to saddle- bifurcations.  All the graphs, for different values of $J$, are zooms made to underline the existence of a region surrounded by the Hopf bifurcation (blue line) on the one side and saddle node bifurcations (red lines) on the others. Such a region is crucial as it describes a regime of bistability: all the selected values of $J$ display a small bistable region, whose size decreases with $J$. } \label{j-diagram}
  \par\end{center}
\end{figure}

The manifolds of Hopf and SNIC bifurcations could be drawn together in a three-dimensional space, using $\omega$, $a$ and $\sigma$ as coordinates.  An alternative easy way to visualize them is to make projections into the $(a, \sigma)$ two-dimensional space for different values of $J$ (see Figure \ref{j-diagram}). Since we set $\omega=1$, and in each such projection $J$ is fixed, the Hopf bifurcation is then obtained as $a_H=a_H(\sigma)$, while the saddle node is obtained by solving Eqs.(\ref{oa-saddle}) as a parametric curve, depending on $R$ and $\sigma$.
Figure \ref{j-diagram} shows  that there is a bistability region delimited by a Hopf line and two saddle-node lines for different values of $J$. This region decreases in size as the coupling decreases until it disappears for sufficiently low values of $J$.

\section{Computational analyses and results} \label{sec-avalanches}

\subsection{Phase diagram of the full model} \label{sec-phase-diagram}

As we have seen, deriving all the phases and transitions between them analytically is a difficult task, and thus, one needs to resort to computational analyses. First of all, an adequate order parameter needs to be defined. As discussed above, the usual Kuramoto order parameter, $R=|Z|$, is not a good choice for inhomogeneous oscillators, because $ \left| \langle Z \rangle_t \right | \neq0$ even when they are uncoupled or asynchronous, due to the different amount of time they spend at diverse phase values.

We employed the so-called Shinomoto and Kuramoto (SK) parameter that solves this problem, being able to discriminate between synchronous and asynchronous regimes \cite{SK-1}. In particular, we consider the --computationally more efficient-- variant of such a parameter employed by Lima and Copelli \cite{Copelli-2019}:
\begin{equation}
  S = \sqrt{\langle |Z|^2 \rangle_t - \left | \langle Z \rangle_t \right | ^2.}
  \label{S}
\end{equation} 
to distinguish the synchronized phase ($S \neq 0$) from asynchronous or excitable states ($S=0$). 

The main result of our computational analyses is the phase diagram reported in Figure \ref{phase_diagram} of the main text. The different phases are identified computationally as follows: 
\begin{itemize}
\item A non-vanishing $S$ value characterizes the synchronous region.
\item The asynchronous and ``collective excitability'' regions are both characterized by $S=0$. To
  detect this second regime one needs to perturb the system and analyze its collective response (or lack of it). 
  
\item The regime of bistability is challenging to study numerically since in finite networks, fluctuations can drive the system to jump between the two coexisting states. To determine this region, the analytical solution was computed by solving for the first $50$ terms in the series expansion (\ref{zk_system}). It was then computationally verified that, for large enough network sizes, two alternative stable states exist within such a region.
\end{itemize}

\subsection{Accuracy of different closures}
For completeness, we checked the accuracy of the different considered closures by comparing them with the results of computational analyses. In order to do so, the different ansatzes or closures:  (i) the Ott-Antonsen (eqs. (\ref{OA})), (ii) the Gaussian closure (eqs. (\ref{gaussian})),  and (iii) the equations proposed by Tyulkina \emph{et al.} (eqs. \ref{tyulkina}) were solved near the Hopf and SNIC bifurcation branches and compared the results --shown in Figure \ref{closures-vs-sim}-- with those of direct simulations of Eq.(\ref{SNIC-coupled}) as reported above. We would like to remark that, up to our knowledge, it is the first time that the accuracy of the different closures to locate different kind of bifurcations has been formally checked. 

\begin{figure}[hbtp]
\begin{center}
  \includegraphics[width=1.0\columnwidth]{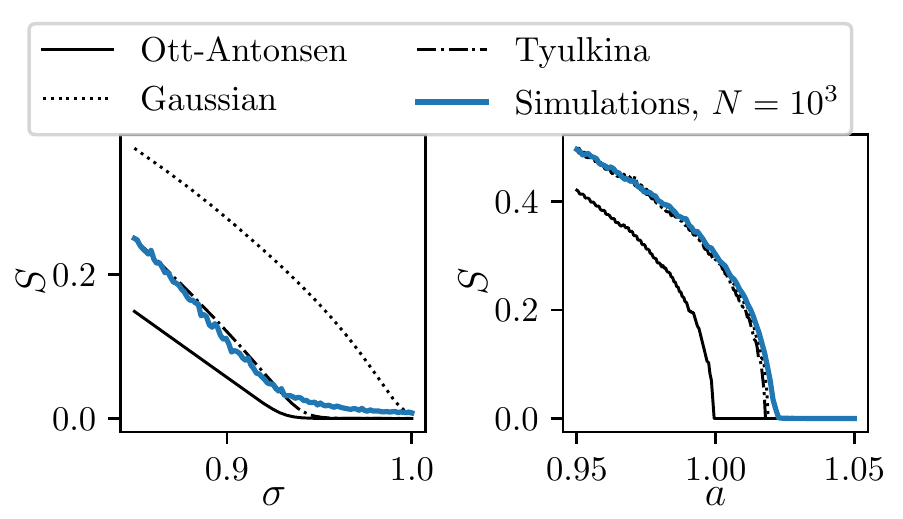}
  \caption{\textbf{Comparison between the results obtained for analytical closures and direct simulations.} Shinomoto-Kuramoto parameter $S$ along the Hopf and SNIC bifurcations, as measured both in simulations and numerical solutions obtained from different closures (as specified in the legend). Parameters: $\omega=J=1$. Hopf, $a=0.5$. SNIC, $\sigma=0.275$. Direct simulations are performed in a fully connected network of size $N=10^3$ (see Section \ref{avalanchesSec} for computational details.)} \label{closures-vs-sim}
  \par\end{center}
\end{figure}

The conclusion is that the ansatz by Tyulkina \emph{et al.} \cite{Pikovsky-reduction} fits more accurately both the Hopf and SNIC branches than the other ones. The Gaussian ansatz captures very well the phenomenology near the SNIC bifurcation but not near the Hopf one. On the other hand, the Ott-Antonsen solution does not predict the transition accurately at any of the bifurcations. However, the ansatz by Tyulkina \emph{et al.} \cite{Pikovsky-reduction} is not able to predict the existence of the bistability phase, while the other two do so. Solving the complete system of equations for the Kuramoto-Daido parameters, at least 5 harmonics are needed in order to find the bistability region.

\subsection{Avalanches at different types of bifurcations} \label{avalanchesSec}
Here, we further investigate whether scale-free avalanches emerged when we moved away from the bistability region.  Results at the different type of bifurcations are reported in Figure 2 of the main text, which displays raster plots computed at either a (Type II) Hopf (upper panels) or at a (Type I) SNIC bifurcation point (lower panels), respectively, as well as within the synchronous and asynchronous phases surrounding them.
Here, we show results for values slightly below, at, and above the bifurcation. As discussed at extent in the main text, scale-free avalanches (with power-law distributed sizes and durations) emerge only in the vicinity of the hybrid type synchronization transition; when the system is moved away from it, scaling behavior as well as scale-free behavior breaks down (see  Figure \ref{break}), while near the bistability region scale-free distributed avalanches remain (see  Figure \ref{Bist}).

\begin{figure}[hbtp]
\begin{center}
  \includegraphics[width=1.0\columnwidth]{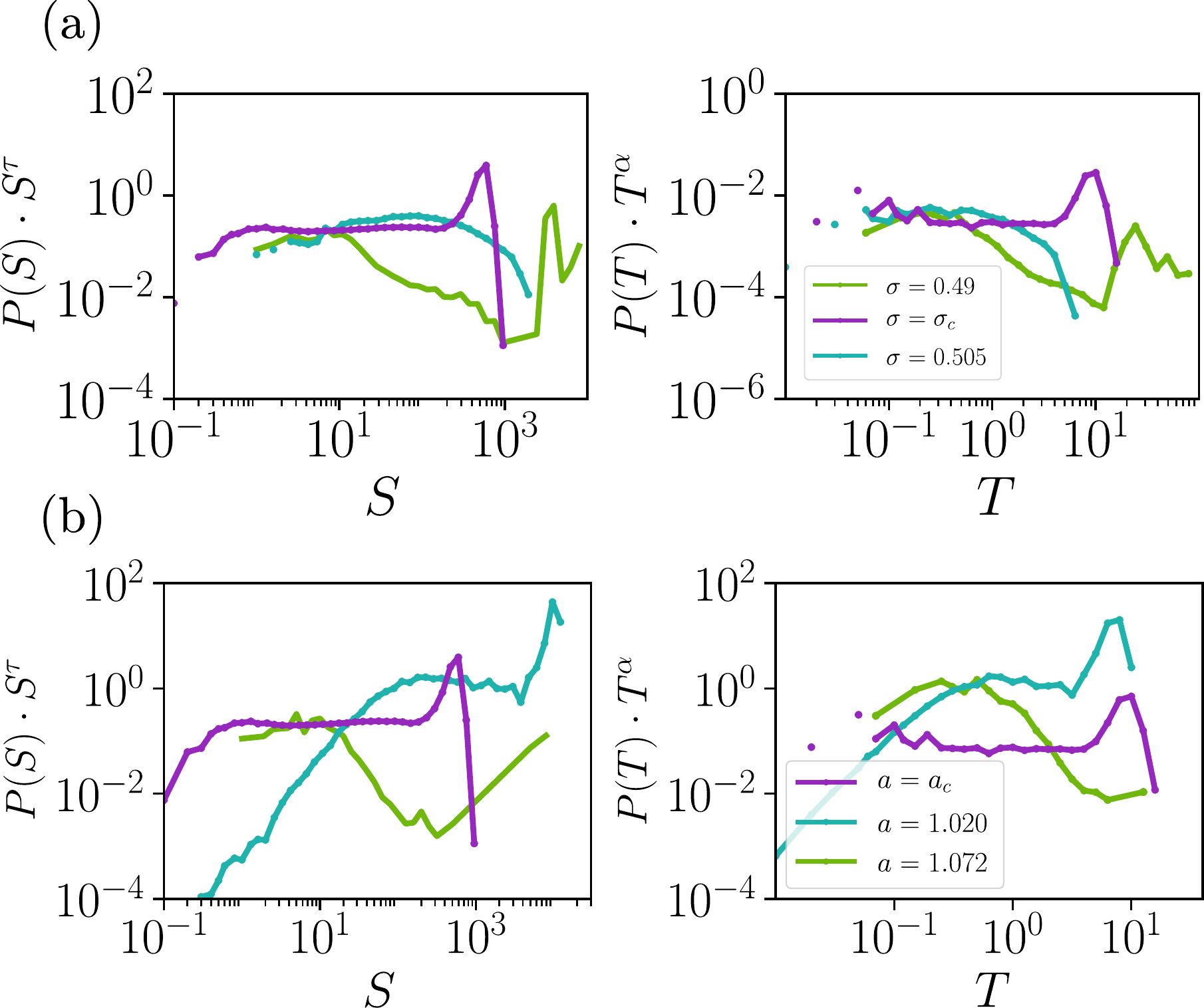}
  \caption{\textbf{Avalanches at and away from the hybrid type synchronization transition.} (a) Avalanche-size and avalanche-duration distributions for a network of size $N=5000$ evaluated at the
hybrid type synchronization transition ($a=1.07$, $\sigma_c=0.496$)
 and two other nearby points,  slightly away from it. The Figure includes: (a) avalanche-size a duration distributions for several values of the noise intensity $\sigma$ (see legend) with fixed $a_c=1.07$ and (b) avalanche-size and duration distributions  for several values of the excitability $a$ (see legend) keeping $\sigma_c=0.496$ fixed. Power-law behavior is observed only at the hybrid type transition.}
     \label{break}
\end{center}
\end{figure}

\begin{figure}[hbtp]
\begin{center}
  \includegraphics[width=1.0\columnwidth]{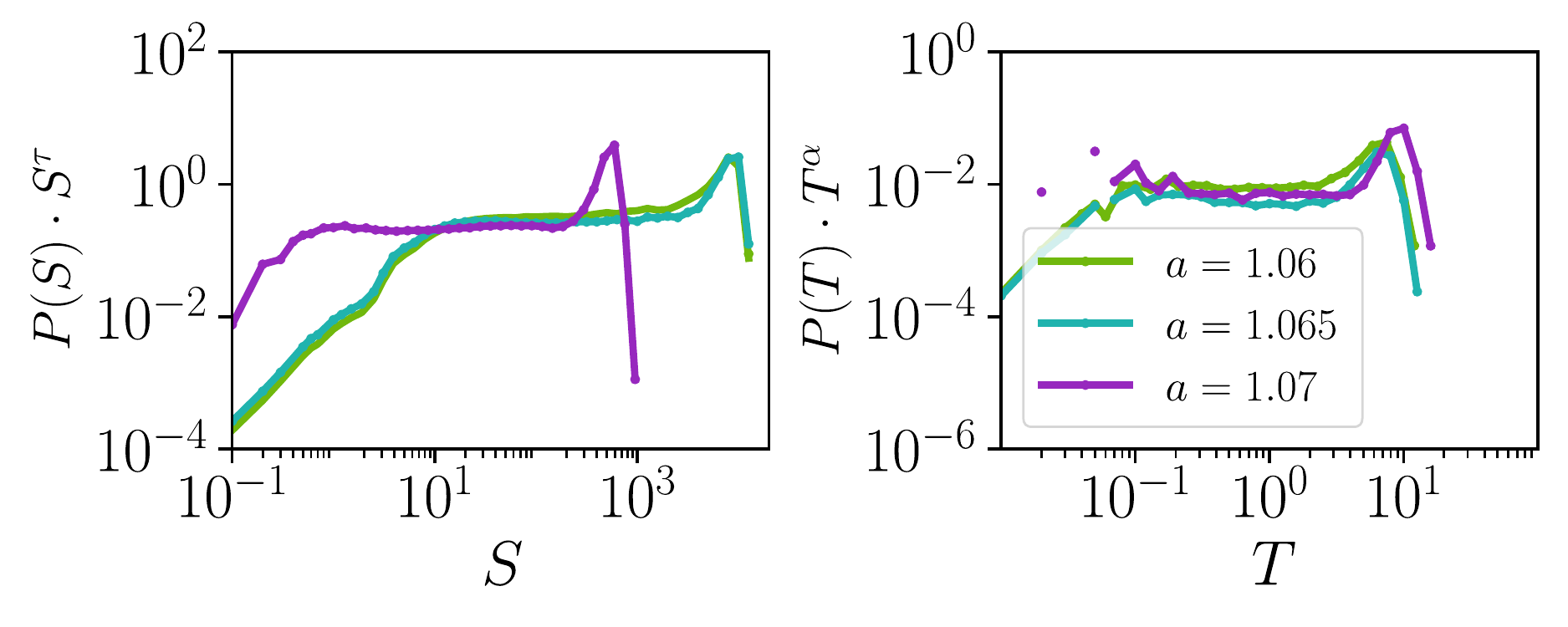}
  \caption{\textbf{Avalanches close to the hybrid type synchronization transition} Avalanche-size and avalanche-duration distributions for a network of size $N=5000$ evaluated at the
hybrid type synchronization transition ($a=1.07$, $\sigma_c=0.496$)
 and two other nearby points, slightly to the left of it. Power law behavior is observed only in the bistable region close to the hybrid-type synchronization.}
     \label{Bist}
\end{center}
\end{figure}

\subsection{Dynamical variability}

As a complementary measure of complexity at the different transition points,  we have also computed the probability distribution of the inter-spike intervals (ISI), along with its associated coefficient of variation (CV) 
\begin{equation}
\cv = \left\langle \sigma_{\isi}/\mu_{\isi} \right\rangle
\end{equation}

where $\mu_{\isi}$ and $\sigma_{\isi}$ are the mean and standard deviation of the inter-spike interval for each oscillator, respectively, and $\langle\cdot\rangle$ indicates an average over units. It is important to distinguish between the interspike interval of each single unit, and the interval between any two consecutive spikes in a network. When looking at avalanches, the second option is preferred, since it gives a small timescale able to resolve the internal structure of avalanches. On the other hand, to measure the CV one focuses on inter-spike intervals of individual oscillators.

For a Poisson process, one expects an exponential distribution of $\isi$'s along with a $\cv=1$; as a rule of thumb, $\cv > 1$ is the fingerprint of irregular spiking activity.  Figure \ref{spikes_stats} shows the probability distribution of the ISIs, and the CVs for the different phases and transitions, as determined in computational analyses.
Only two cases exhibit $\cv > 1$: the hybrid type synchronization transition, as well as a small neighborhood around it in the bistable regime; they are also the only two cases characterized by a broad distribution of ISI values. Thus, a high level of variability --similar to that observed in the cortex-- is only found in the region around the hybrid type synchronization transition, but not in  the neighborhood of either standard Hopf or SNIC bifurcations.

\begin{figure}[hbtp]
\begin{center}
 \includegraphics[width=1.0\columnwidth]{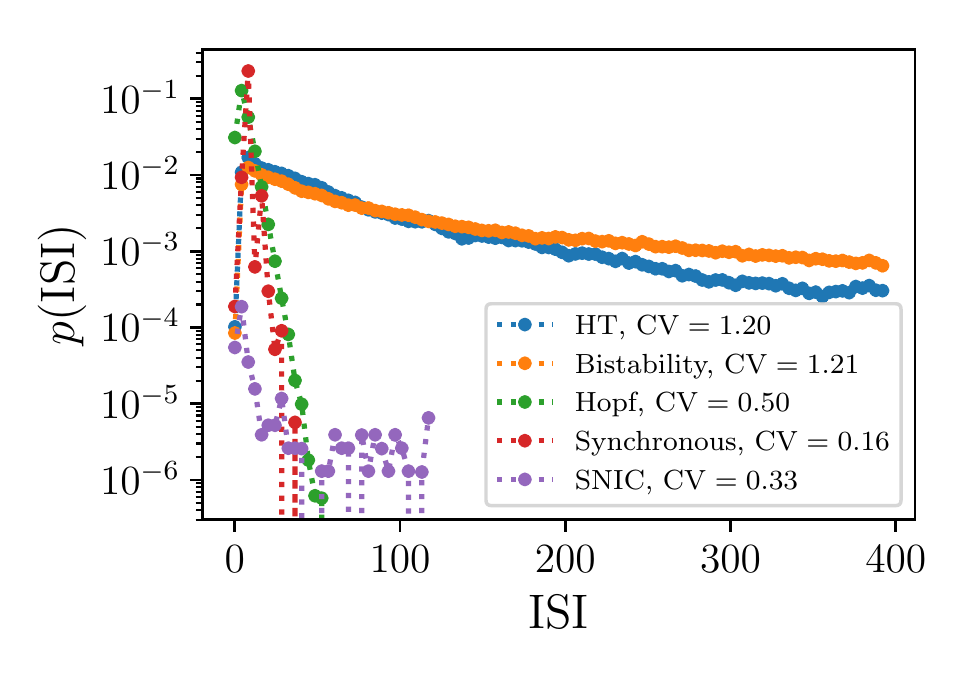}
 \caption{\textbf{ISI distributions and CVs for fully-connected networks. }  Probability distributions for inter-spike intervals (ISI) for different phases and bifurcations as shown in the legend. The coefficients of variation for each case are indicated also in the legend. Parameter values: Synchronous regime, $\sigma=0.5$, $a=0.5$; Hopf bifurcation, $a=0.5$, $\sigma=0.92$; collectively excitable phase, $\sigma=0.5$, $a=1.12$; hybrid type synchronization transition: $\sigma=0.5$, $a=1.07$; bistable regime: $\sigma=0.5$, $a=1.072$. Network size $N=5000$.}
  \label{spikes_stats}
 \par\end{center}
\end{figure}

\section{Phase diagram for the model on two-dimensional lattices}

\begin{figure}[h]
\begin{center}
  \includegraphics[width=1.0\columnwidth]{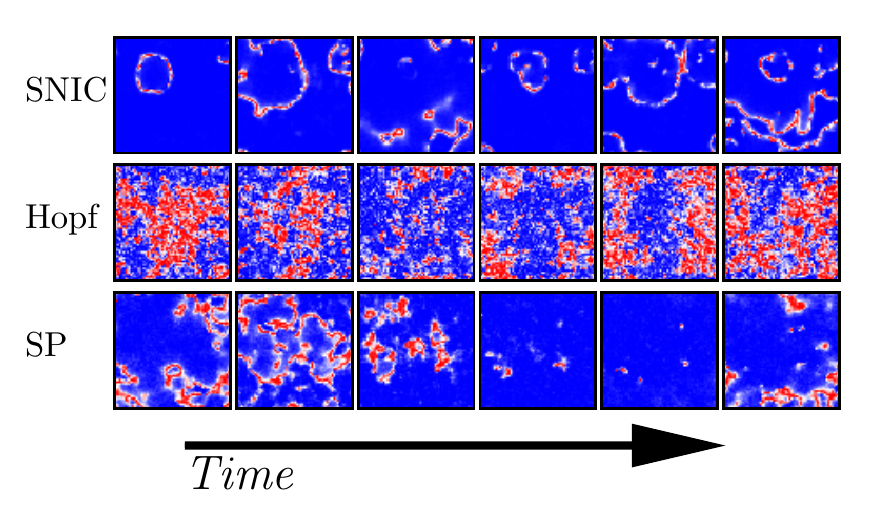}
  \caption{\textbf{Spatio-temporal dynamics on two-dimensional systems}
The figure shows three rows, each one with six different frames corresponding to six different times on a running simulation, for the following cases:
    Upper row:  near the different bifurcations SNIC ($a=1$, $\sigma=0.08$); central row: near a Hopf bifurcation ($a=0.5$, $\sigma=0.65$);  and lower row: hybrid type synchronization transition ($a=0.98$, $\sigma=0.185$). Blue color indicates lack of activity, while red color stands for maximum activity levels (identified as $1+\sin\phi_j$). Near the SNIC transition, noise fluctuations generate wavefronts that propagate in the system. At the Hopf transition, there is some background activity whose level grows and shrinks periodically.
    At the hybrid type synchronization transition, the spatio-temporal patterns are more complex, being a mixture of the two aforementioned types; in particular, the system sometimes falls in the excitable state. Simulations performed for $N=64^2$ with periodic boundary conditions. Videos showing the evolution for both cases and the other phases can be found in the Supplementary Material 1 \cite{SI}.}\label{2D-frames}
  \par\end{center}
\end{figure}

Computational analyses in 2D systems reveal a very similar phase diagram to the mean-field one but with a richer phenomenology (as graphically illustrated in Figure \ref{2D-frames}). There are three main phases, in a nutshell: a synchronous regime, an asynchronous one, and a collectively excitable phase, much as in the mean-field case.
In the asynchronous regime, clusters of activity appear, propagate, and vanish dynamically, with an averaged constant \new{network activation level} (smaller/larger close to the SNIC/Hopf transition, respectively) but with no overall synchronization. On the other hand, within the synchronous phase, activity wane and washes and periods of overall quiescence are followed by bursts of overall activity that spreads quickly from different focuses.
Such a collective propagation requires some level of synchronization (let us recall that a perfectly phase-synchronous state does not exist in 2D \cite{SK-2}  as rotational symmetry cannot possibly be broken in low-dimensional systems; there are always some ``topological defects'' in the system preventing it to exhibit perfect synchronization, as it happens in well-known models of equilibrium statistical mechanics \cite{MerminWagner} and also in the Kuramoto model \cite{Acebron}).
Finally, the excitable state has all units in the ``down'' regime, with almost no activity, but is susceptible to respond to external perturbations or inputs.

Let us discuss the observed phenomenology at the different bifurcation lines separating these phases:

(i) Nearby the type-I synchronization transition, at the SNIC bifurcation (slightly within the synchronous phase), \new{oscillating activity appears in the form of traveling waves}. Visual inspection reveals, e.g. the presence of typical spiral patterns typical of two-dimensional excitable systems (\newr{see Supplementary Material 1 \cite{SI} for videos, and Figure \ref{2D-frames} upper row)}.

(ii) Nearby the type-II synchronization transition, at the Hopf bifurcation (slightly within the synchronous phase) the overall level of activity oscillates in time (i.e., the system ``breathes'')  and is spatially distributed in fragmented clusters (see Figure\ref{2D-frames}, central row).

(iii) Near the hybrid type synchronization transition, the dynamical behavior is much more complex: somehow in between the overall oscillatory behavior along the Hopf line and the emergence of wavefronts at the SNIC line (see Figure \ref{2D-frames} bottom row).

Thus, even if we are not attempting to quantify spatio-temporal complexity here, it is clear that more complex dynamics emerge around the HT synchronization transition.
As shown in Figure \ref{Diagram2D} the computationally-obtained phase diagram in the case of two-dimensional lattices has a structure qualitatively identical to the mean-field case, including the same phases.

\begin{figure}[h]
\begin{center}
  \includegraphics[width=1.0\columnwidth]{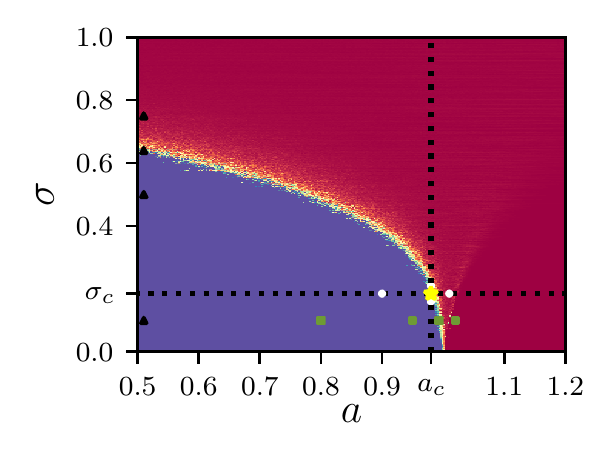}
  \caption{\textbf{Phase diagram for a 2D lattice of coupled excitable oscillators} for different values of $\sigma$ and $a$, using the Shinomoto-Kuramoto (SK) order parameter (color-coded). Reddish colors stand for asynchronous states, while blueish colors correspond to partially synchronized ones. As in the FC-network case, the system can de-synchronize either through a supercritical Hopf bifurcation or through a SNIC bifurcation. In an intermediate region it is possible to find a hybrid type synchronization transition (yellow star). This is characterized as the point where  power-law distributed avalanches obeying finite-size scaling appear. \new{Symbols describe the coordinates for which the videos of Supplementary Material 1 \cite{SI} were recorded:} ``Hopf'' bifurcation (black triangles); ``SNIC'' bifurcation  (green squares) and the HT transition point (white circles and yellow star).}\label{Diagram2D}
 \end{center}
\end{figure}


%

\end{document}